\documentclass{optica-article}
\journal{opticajournal} 

\articletype{Research Article}

\usepackage{placeins} 
\usepackage[separate-uncertainty=true,
            locale = UK, separate-uncertainty, group-separator = {,},
            exponent-product = \cdot
            ]{siunitx}
\usepackage{lineno}
\usepackage[normalem]{ulem}
\usepackage{anyfontsize}

\graphicspath{ {./figures/} }

\begin{document}

\title{Post-compression of a Q-switched laser in a glass-rod multi-pass cell} 

\author{
    Peer Biesterfeld\authormark{1,†,*}, 
    Arthur Schönberg\authormark{2,†}, 
    Marc Seitz\authormark{3,†}, 
    Nayla Jimenez\authormark{2,4,5}, 
    Tino Lang\authormark{2}, 
    Marcus Seidel\authormark{2,4,5}, 
    Prannay Balla\authormark{2}, 
    Lutz Winkelmann\authormark{2}, 
    Jyothish K. Sunny\authormark{1}, 
    Sven Fröhlich\authormark{1}, 
    Philip Mosel\authormark{1}, 
    Ingmar Hartl\authormark{2}, 
    Francesca Calegari\authormark{3,6,7}, 
    Uwe Morgner\authormark{1,8}, 
    Milutin Kovacev\authormark{1,8}, 
    Christoph M. Heyl\authormark{2,4,5,‡} and 
    Andrea Trabattoni\authormark{1,3,8,‡}}

\address{
    \authormark{1}Leibniz University Hannover, Institute of Quantum Optics, Welfengarten 1, 30167 Hannover, Germany\\
    \authormark{2}Deutsches Elektronen-Synchrotron DESY, Notkestraße 85, 22607 Hamburg, Germany\\
    \authormark{3}Center for Free-Electron Laser Science CFEL, Deutsches Elektronen-Synchrotron DESY, Notkestraße 85, 22607 Hamburg, Germany\\
    \authormark{4}Helmholtz-Institute Jena, Fröbelstieg 3, 07743 Jena, Germany\\
    \authormark{5}GSI Helmholtzzentrum für Schwerionenforschung GmbH, Planckstraße 1, 64291 Darmstadt, Germany\\
    \authormark{6}Department of Physics, Universität Hamburg, 22761 Hamburg, Germany\\
    \authormark{7}The Hamburg Centre for Ultrafast Imaging, Universität Hamburg, 22607 Hamburg, Germany\\
    \authormark{8}Cluster of Excellence PhoenixD (Photonics, Optics, and Engineering-Innovation Across Disciplines), 30167 Hannover, Germany\\
    \vspace{2mm}
    \authormark{†}These authors contributed equally to this work.\\
    \authormark{‡}These authors contributed equally to this work.
    \vspace{2mm}
}

\email{\authormark{*}biesterfeld@iqo.uni-hannover.de}

\begin{abstract*} 
Q-switched lasers are compact, cost-effective, and highly pulse energy-scalable sources for nanosecond-scale laser pulses. The technology has been developed for many decades and is widely used in scientific, industrial and medical applications. However, their inherently narrow bandwidth imposes a lower limit on pulse duration - typically in the few-hundred-picosecond range - limiting the applicability of Q-switched technology in fields that require ultrafast laser pulses in the few-picosecond or femtosecond regime.
In contrast, mode-locked lasers can produce broad-band, ultrafast (< \SI{1}{ps}) pulses, but are complex, expensive, and typically require a large footprint. 
To bridge the parameter gap between these two laser platforms - in terms of pulse duration and achievable peak power - we here propose a Herriott-type multi-pass cell (MPC) based post-compression scheme for shortening the pulse durations of Q-switched lasers down to the ultrafast, picosecond regime.
We experimentally demonstrate post-compression of \SI{0.5}{ns}, \SI{1}{mJ} pulses from a Q-switched laser to \SI{24}{ps} using a compact glass-rod MPC for spectral broadening.
We verify this result numerically and show that compression down to a few picoseconds is possible using the nanosecond MPC (nMPC). Through spectral filtering approaches, the nMPC suppresses detrimental nonlinear processes such as stimulated Raman scattering, which have set severe limitations for fiber-based post-compression of Q-switched lasers until today. Our results pave the way to cost-efficient and compact ultrafast laser platforms based on Q-switched laser technology.
\end{abstract*}

\FloatBarrier
\section*{Introduction}
The continuous development of lasers has had a profound impact on our society. In particular, laser pulses with durations in the nano- ($1\text{ ns} = 10^{-9}\text{ s}$) to femtosecond ($1\text{ fs} = 10^{-15}\text{ s}$) range are routinely employed for microsurgery \cite{RefrEyeSurg,Medicine} optical gas sensing \cite{GasSensing}, high-precision surface and volume material processing \cite{MaterProces}, or imaging of biological samples \cite{NeurImage}. Moreover, short-pulsed lasers are playing an indispensable role in fundamental research with applications ranging from the ultrafast spectroscopy of photo-induced electron dynamics in matter \cite{ChargeMigr} via the manipulation of transient states in materials \cite{Superconductivity} to recent breakthroughs in nuclear spectroscopy \cite{Thorium229(1), Thorium229(2)}. Depending on the laser parameters, in particular the required peak power and pulse duration, different pulsed laser technologies are used, employing predominantly Q-switching or mode-locking schemes. \\

On the one hand, nanosecond pulses are typically generated using Q-switching methods \cite{QSwitching, QSwitchedLaser, QSwitchSatAbs, PulsedLaserGener, HighEQSwitch}. Q-switched lasers can support high pulse energy (millijoule to joule), they are cost-efficient and can be built as robust and compact modules. Their peak power, however, is limited, hindering their effectiveness, for example, in material machining, dermatological treatment, or as driving light sources for nonlinear optical processes. On the other hand, shorter durations (pico- to femtoseconds) are the realm of mode-locked lasers \cite{PulsedLaserGener, ModeLockUltra, RevModeLock, RevModeLockFiber}. Such light sources provide pulses that enable the analysis of ultrafast phenomena at femtosecond time scales and furthermore support ultrahigh peak powers, which make them ideal for machining applications, at the expense, however, of high complexity, large footprints and much higher costs. In this context, the idea of compressing the duration of nano- to picosecond or sub-picosecond laser pulses is particularly appealing, as it offers a route to bridge the gap - at least in terms of pulse duration and achievable peak power - between Q-switching and mode-locking laser technologies. \\

\begin{figure}[ht!]
    \centering
    \captionsetup{width=\linewidth}
    \includegraphics[width=\textwidth]{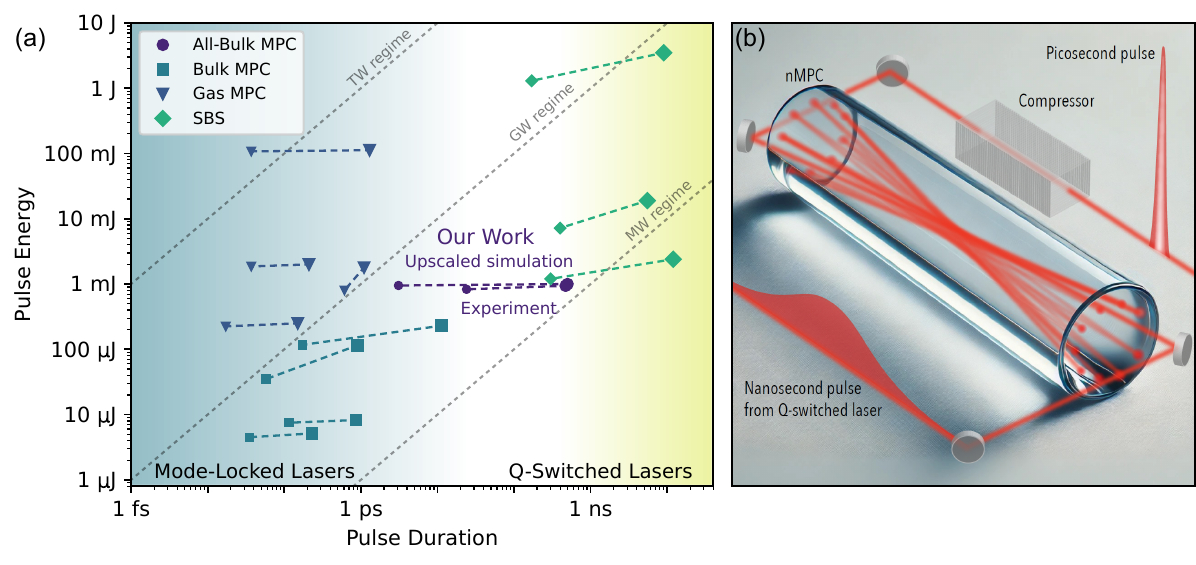}
    \caption{\textbf{Post compression methods covering the parameter space from nanoseconds to femtoseconds.} (a)
    Parameter overview of representative laser pulse post-compression works. Q-switched laser systems (green) have been post-compressed via SBS (diamonds), while gas-filled (triangles) or bulk (squares) MPCs are widely used for compressing mode-locked laser systems (blue) \cite{Sub40fsMPC, Weitenberg2017, Gen172fsMPC, FLASHpsMPC, DispEngMPC, HighEGasMPC, FELMPC, LG10MPC, Gen300psSBS, SubnsSBSMPC, HighESBS}. Our work (purple) bridges the gap between these two laser regimes via post-compression of Q-switched lasers in a Herriott-type, bulk-rod MPC. The experimental output data point takes into account the output pulse duration measured at reduced energy, scaled to the full energy transmitted through the MPC. (b) Illustration of the proposed nanosecond MPC concept.
    }
    \label{fig:overview}
\end{figure}

In the last four decades, post-compression methods, i.e., technologies employing nonlinear optical processes to spectrally broaden and temporally compress the laser pulse, have tackled this challenge. Self-phase modulation (SPM) in hollow-core capillaries/fibers \cite{PostCompressionTech, AdvancesHCF, Nisoli1996}, photonic crystal fibers \cite{PCFsub70fs, PCFsub2Cycle, PCFKagome}, or multi-pass cells (MPCs) \cite{Schulte2016, Weitenberg2017, MPCReview, TheoryMPC} is used to compress mode-locked lasers even down to the few-femtosecond range. 
However, the compression of Q-switched lasers has proven more challenging. Compression methods exploiting stimulated Brillouin scattering (SBS) in bulk or liquid media or by propagation in long (> 100 m) optical fibers, enabled to compress nanosecond lasers into the picosecond regime \cite{SubnsSBSMPC, HighESBS, IntLight, Johnson1984, FiberComprSoliton}. 
However, these efforts were limited in performance by phonon lifetime constraints or competing nonlinear effects - for instance, stimulated Raman scattering (SRS) - as well as in energy scalability, especially for the fiber based approaches.
As a result, post-compressing Q-switched lasers to the parameter space of mode-locked lasers at millijoule-class pulse energies has not been achieved until now. %
Fig. \ref{fig:overview}(a) illustrates this discussion, showing a representative set of post-compressed Q-switched and mode-locked lasers. The figure visualizes the input laser parameters (larger shapes) and the post-compressed output (smaller shapes), connected with dashed lines. 
The distribution of the displayed data points underlines a major parameter gap between the two laser technologies. \\

Here, we introduce a novel technology based on a bulk-rod Herriot-type multi-pass cell, that opens a route to compress lasers with few MW or sub-MW peak power into the pico- to femtosecond regime, approaching GW power levels. We experimentally demonstrate the technique by compressing a 0.5\,ns pulse to 24\,ps using a 10-cm-long fused silica (FS) rod as MPC nonlinear medium. Our simulations demonstrate that the method can scale to compression factors exceeding 100. 
Spectral filtering thereby prevents the rise of detrimental nonlinear processes including stimulated Raman scattering and quasi-phase-matched four-wave-mixing.  
Our approach opens a perspective towards low-cost, compact, high-peak-power laser systems for novel implementations in science and industry. This includes, for example, the generation of high-flux extreme-ultraviolet light, high-precision laser-driven material machining, or damage-free tissue treatment in dermatology. 

\FloatBarrier
\section*{Concept}
\label{sec:concept}

Compressing the temporal duration of laser pulses relies on the concept of spectral broadening. Nonlinear optical processes such as self-phase modulation (SPM) are employed for this purpose, being implemented in a variety of post-compression schemes \cite{MPCReview}. However, compressing q-switched pulses with nanoseconds duration presents a main challenge: SPM requires sufficient peak power (usually ranging from 10s of MW to the GW level) as well as long propagation lengths inside a nonlinear medium in order to accumulate nonlinear phase. However, this peak power regime is usually not reached by millijoule-class q-switched lasers. To the best of our knowledge, pulses with a peak power down to about \SI{8.7}{MW} have been post-compressed through SPM with an MPC \cite{MPCReview}. For comparison, a typical q-switched laser with a pulse energy of 1\,mJ and a duration of 1\,ns provides a peak power of only 1\,MW. Long propagation in solid-core fibers with high optical nonlinearity may appear as a good mitigation strategy and has been employed to successfully post-compress nanosecond pulses in the nanojoule-energy regime \cite{Kafka1984, Johnson1984}. However, Raman and/or Brillouin scattering typically deteriorate the spectral broadening process, limiting energy-scalability especially for nanosecond pulses. \\

In order to overcome the above-mentioned limitations, we introduce a new type of multi-pass cell made completely out of glass such as fused silica. The monolithic MPC is equipped with high-reflectivity (HR) coated, curved surfaces acting as cavity mirrors (Fig. \ref{fig:overview}(b)). This type of MPC, dubbed nMPC (nanosecond MPC), presents unique advantages for long (nanosecond to hundreds of picoseconds) driving pulses. First, it provides both large nonlinear coefficients as well as long nonlinear interaction lengths while keeping the overall footprint of the setup compact.
Second, while input pulses with pico- or femtosecond duration would suffer from the high dispersion in the bulk material, nanosecond pulses are insensitive to dispersion due to their narrow bandwidth. 
Third, the laser-induced damage threshold (LIDT) of the optical coating of the curved surfaces is much higher for nanosecond lasers than for shorter pulses, with values in the range of 10s of \SI{}{J/cm^2} \cite{Niemz1995, Atkocaitis2022}. Consequently, MPC configurations with small beam sizes on the mirrors can be realized, thus allowing for many round trips and large spectral broadening factors. 
Fourth, the optical coatings can act as a spectral filter at each round trip, suppressing competing nonlinear optical processes such as Raman scattering and four-wave mixing that otherwise would strongly limit the pulse compressibility (further details in the discussion). 

\begin{figure}[ht!]
    \centering
    \captionsetup{width=\linewidth}
    \includegraphics[width=\textwidth]{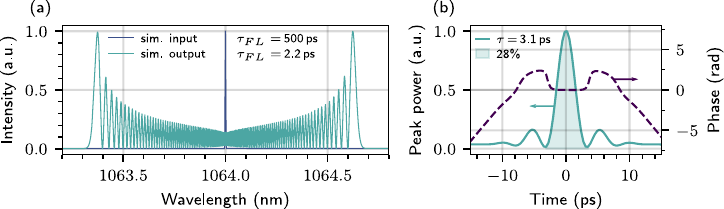}
    \caption{\textbf{Simulated post-compression using the nMPC concept.} (a) Input and output spectra demonstrating a high compression rate in the proposed full-bulk nMPC. (b) Corresponding compressed pulse, where the phase is compensated using GDD only.}
    \label{fig:outlook}
\end{figure}

Based on this concept, simulations with an input pulse energy of \SI{1}{\m J}, an input pulse duration of \SI{0.5}{ns} of a Gaussian laser pulse and \SI{150}{} round-trips (N=150, k=91), are carried out to demonstrate the potential of the nMPC. The simulation is performed using a symmetric split-step Fourier algorithm to solve the Unidirectional Pulse Propagation Equation in (2+1) dimensions, assuming radial symmetry \cite{Couairon2011}. 
We consider a FS glass block with curved surfaces of $R=\SI{10}{cm}$, length $L=\SI{25.43}{cm}$ and an MPC configuration parameter $k=91$ \cite{MPCReview}. The in/out coupling can be achieved through a small uncoated spot in one cell surface. 
By using a diameter of 3 inch (\SI{7.62}{cm}) for the curved surfaces, the necessary 150 reflections can be fit onto the curved surfaces.
As an alternative route, a more dense round-trip pattern can be employed, overcoming geometric limitations of a simple circular pattern \cite{Robert2007}.
Figure \ref{fig:outlook} shows the results of the simulation.
The spectral bandwidth of the input $\Delta\lambda_{10\text{dB}}=\SI{6}{pm}$ increases by a factor of 220 to $\Delta\lambda_{10\text{dB}}=\SI{1.32}{nm}$, supporting a pulse duration of \SI{2.2}{ps} (Fig. \ref{fig:outlook}(a)).
Following second-order phase compensation (\SI{103}{ps\squared}), a compressed pulse with a duration of \SI{3.1}{ps} is obtained, showing very good compressibility. This results in a temporal compression by a factor of \SI{161}{} and a final peak power of \SI{82}{MW} (Fig. \ref{fig:outlook}(b)). 

In the all-bulk scheme, the main sources of loss are the coated end facets of the bulk material and the absorption in the medium. With modern manufacturing methods, these losses can be kept very low, allowing the realization of such an MPC with an efficiency \SI{>90}{}\%. Considering commercially available transmission gratings with efficiencies of \SI{>98}{}\% the total transmission of the setup can be in the order of \SI{90}{}\%. Alternatively, slightly less efficient but very compact compression can be achieved using a dispersive volume Bragg grating. The efficiency in the required parameter regime is $\gtrsim\SI{85}{}\%$\cite{Glebov2014}. This allows for a total post-compression efficiency of \SI{>75}{}\% while maintaining a very small footprint.

\FloatBarrier
\section*{Experimental demonstration}
\label{sec:experiment}

\begin{figure}[ht!]
    \centering
    \captionsetup{width=\linewidth}
    \includegraphics[width=.95\textwidth]{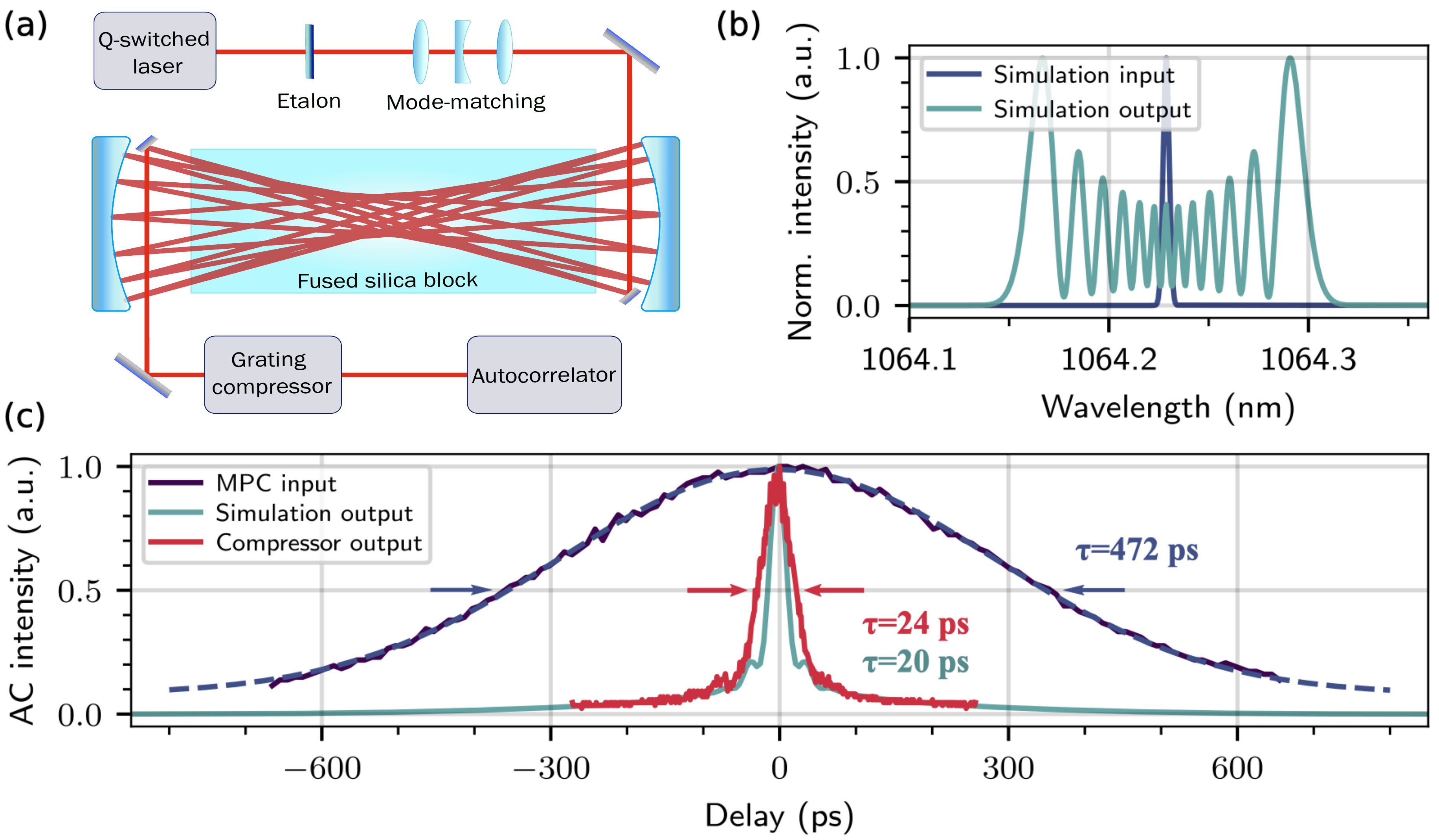}
    \caption{\textbf{Experimental setup and results.} (a) Schematic experimental setup including in particular an MPC using a \SI{10}{cm} FS rod as a nonlinear medium. (b) Spectral broadening characteristics obtained via numerical (2+1)D simulation mimicking the experimental conditions. (c) Autocorrelation (AC) of the MPC input and compressor output pulse, along with the AC of the GDD-compensated simulated output. The pulse durations $\tau$ are obtained by fitting a Gaussian (input) and a Lorentzian (output) to the experimental data, taking the corresponding deconvolution factors into account.}
    \label{fig:comp_results}
\end{figure}

We experimentally demonstrate the nMPC concept using a simplified setup which is schematically depicted in Fig. \ref{fig:comp_results}(a)). We employ a commercial Q-switched diode-pumped Nd:YAG laser (MPL2310, QS Lasers), emitting at \SI{1064}{nm}. The passive Q-switching is achieved by a Cr:YAG crystal, resulting in the generation of pulses with a duration (full-width at half maximum, FWHM) of \SI{0.4}{ns}. With an ultracompact footprint ($\SI{10}{}\times\SI{17}{cm}$), the laser features a pulse energy of \SI{2}{mJ} at a repetition rate of \SI{100}{Hz}. The laser cavity is operated in single-mode emission upon thermal stabilisation, yielding only (\SI{<0.1}{}\%) of energy in adjacent longitudinal modes. The single-mode contrast is further enhanced by using an Etalon filter with a free spectral range of \SI{230}{pm} and a finesse of 41, resulting in an additional reduction of \SI{24}{dB} for the adjacent modes, leading to a mode contrast of $< \SI{0.001}{}\%$. The effect of multi-mode operation on the performance of the nMPC is discussed in detail in the Methods section.

For the sake of simplicity of this first test setup, we decomposed the monolithic bulk-rod nMPC into a set of two cavity mirrors, a large glass block, and an in-coupling scraper mirror. We use a $\SI{10}{cm}$ long block of anti-reflection coated, UV-FS as nonlinear medium, placed in the center between two concave mirrors with a radius of curvature of \SI{0.1}{m} (Fig. \ref{fig:comp_results}(b)). 
The input beam is mode-matched to the MPC by a 3-lens telescope and is coupled into the cell by the scraper mirror with a width of \SI{4}{mm}. The laser pulse undergoes 62 passes in the nMPC. Specifically, the MPC is configured with N=31 and k=24, where N defines the number of round trips and k specifies the chosen configuration which fulfills the re-entrance condition \cite{MPCReview}. In this configuration, the linearly mode-matched MPC mode corresponds to a fluence of \SI{0.3}{J/cm\squared} on the cell mirrors and a peak intensity of \SI{1e10}{W/cm\squared} at the focus. LIDT measurements have been conducted on the glass block, showing laser-induced damage inside the bulk for a driving peak intensity of \SI{2e10}{W/cm\squared}.  The in-coupled pulses have an energy of \SI{0.94}{mJ} and a duration of \SI{0.47}{ns} (FWHM), measured with an in-house developed second-order autocorrelator. The MPC transmission efficiency is measured to be \SI{83}{}\%. This value is in good agreement with the calculated efficiency of \SI{86}{}\%, considering the reflectivity of the cell mirrors and the FS bulk surfaces as specified by the manufacturer. The beam spatial quality is measured before and after the nMPC, showing a slight degradation of the mean M-square from 1.31 to 1.45 (see Figure \ref{fig:beam_results}). 

Figure \ref{fig:comp_results}(b) shows the simulated SPM-driven spectral broadening occurring in the nMPC for the current experimental parameters. Starting from an input bandwidth of around \SI{6}{pm}, an output bandwidth of \SI{155}{pm} is obtained through SPM, corresponding to a transform-limited duration of \SI{20}{ps} (FWHM). The spectral phase of the simulation output is compensated via GDD compensation ($\SI{-1370}{ps^2}$), resulting in a pulse duration close to the Fourier-limited value and an energy content of \SI{60}{}\% in the main pulse.  Experimentally, spectral broadening is monitored using an Echelle spectrometer with a resolution of \SI{35}{pm} at \SI{1064}{nm}. Although this instrument is sufficient to show a signature of SPM, it is not able to resolve the broadened spectrum. For this reason, only the simulated spectra are reported in Fig. \ref{fig:comp_results}(b), while the experimental characterisation of the input and output pulses was performed in the time domain. 
Following spectral broadening in the nMPC, the output beam is collimated and sent to a transmission-grating compressor. The compressor consists of two gratings with a line density of \SI{1740}{l/mm}, resulting in a total chirp of approx. \SI{-1100}{\ps\squared} and a limited efficiency of $\eta = \SI{12}{}\%$). We note that this compressor was used simply for availability reasons to demonstrate compressibility, disregarding efficiency. The used gratings can easily be replaced by tailored transmission gratings or by a chirped volume Bragg grating (CVBG). A CVBG can provide the required chirp while offering high efficiency \SI{\sim90}{}\%, supporting high pulse energies and average powers exceeding \SI{1}{mJ} and \SI{250}{W} in a single compact optical element measuring a few centimeters in length \cite{Glebov2014}. 

The  duration of the compressed pulses is measured to be \SI{24}{ps} (FWHM), which is close to the simulated transform limit of \SI{20}{ps} (Fig. \ref{fig:comp_results}(c)). The compression ratio, based on the measured input duration of \SI{0.47}{ns} (FWHM), is approximately \SI{20}{}. This result demonstrates the effective compressibility of single-mode q-switched lasers in SPM-driven MPCs.

\FloatBarrier
\section*{Discussion}
\label{sec:discussion}

Our proof-of-principle experiment demonstrates the nMPC concept and shows the first MPC-based compression of nanosecond-class laser pulses. The compressed pulses which can be obtained in our experiment using an improved compressor setup match the driving peak power range of post-compression setups already demonstrated in literature  \cite{Weitenberg2017}. In addition, compression factor up-scaling to reach shorter output pulses in a single stage using our scheme appears feasible (see Fig. \ref{fig:outlook}).
Combining an nMPC with a conventional compression method thus holds promise to fully bridge the gap between nanosecond and femtosecond pulse durations, opening up a new class of femtosecond lasers. 

Taking into account future parameter scaling approaches, it is important to discuss the accessible parameter space supported by an nMPC. In this context, it is critical to consider that multiple detrimental processes can be triggered upon extreme spectral broadening in bulk media, and their contribution depends on the bandwidth of the input pulse. 
In this framework, we identify five major phenomena: temporal quality of post-compressed pulses, longitudinal mode beating of Q-switched lasers \cite{Corless1997}, stimulated Brillouin scattering (SBS) \cite{Chiao1964, Armandillo1983, Agrawal2013}, stimulated Raman scattering (SRS) \cite{Agrawal2013}, and degenerate quasi-phase-matched four-wave-mixing (QPM-FWM) \cite{Hanna2020}.
We discuss the impact of these processes for post-compression in an nMPC along with suitable mitigation strategies.\\

The degradation of the temporal quality for high post compression factors in a single stage is a well-known challenge \cite{Escoto2022}. This effect is also seen in our simulations: it leads to a low energy content in the main pulse of only \SI{28}{}\% for a compression factor of 161 by considering a Gaussian input pulse shape (Fig. \ref{fig:outlook}).
This effect results from the strongly modulated spectral shape and phase of an SPM broadened Gaussian pulse \cite{Escoto2024}. However, the temporal characteristics for large compression factors can be improved by separation into multiple stages, enhanced frequency chirping or nonlinear polarization ellipse rotation\cite{Escoto2024, Pfaff2022, Benner2023}. 
Another option is shaping the temporal intensity profile of the input pulse into a parabolic shape, which can lead to a high-contrast SPM-post-compressed pulse \cite{Nguyen2011}.
In the nanosecond regime, direct temporal pulse-form shaping technology is readily available down to sub-ns levels with ps resolution \cite{Rogers2016, Meijer2017}. 

\begin{figure}[ht!]
    \centering
    \captionsetup{width=\linewidth}
    \includegraphics[width=\textwidth]{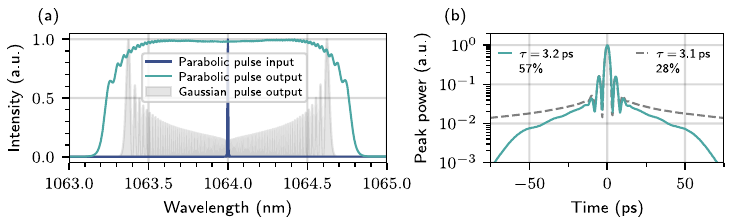}
    \caption{\textbf{Simulated nMPC post-compression using a parabolic temporal pulse shape.} (a) Spectral broadening of a temporally parabolic shaped pulse in comparison to a Gaussian pulse (identical to the data displayed in Fig. \ref{fig:outlook}). (b) Corresponding GDD-compensated pulses, along with the energy fraction contained in the main pulse.}
    \label{fig:parabola_pulse}
\end{figure}

Figure \ref{fig:parabola_pulse} shows this approach in a (2+1)D simulation with an input pulse energy of $E_p=\SI{0.5}{mJ}$, \SI{225}{} round trips through the MPC and a $B$-integral per pass of $B_\text{pass}=\SI{0.2}{}\pi$. The number of round trips $N$ has been chosen to match the Fourier transform limit (FTL) of the broadened spectrum displayed in Fig. \ref{fig:outlook} for direct comparison. All parameters not specified above are kept as in Figure \ref{fig:outlook}. The parabolic input pulse generates a much smoother output spectrum, increasing the energy content of the main pulse by a factor of \SI{2}{}. Our simulations indicate that a further improved energy content in the main pulse can be achieved for decreasing $B_\text{pass}$, compensated by an increased $N$.   

The next challenge for Q-switched laser pulse compression is longitudinal mode beating, which can occur due to additional longitudinal modes emitted from the laser itself, a typical phenomenon for Q-switched lasers \cite{Sooy1965}.
The resulting modulation of the temporal profile of the pulse occurring due to temporal mode interference can distort the phase of the SPM-broadened spectrum, degrading the compressibility of the pulse, as observed in our experiments (further details in the Methods section). Our numerical simulations reveal that an additional mode with an energy fraction of only $10^{-4}$ of the main pulse can already have severe effects on the output spectrum. Thus, mode cleaning before coupling into the nMPC becomes necessary.

Another challenge is Stimulated Brillouin Scattering (SBS), which arises from the nonlinear interaction of an incoming light wave (pump) with the acoustic modes of the medium via electrostriction, generating a periodic modulation of the material density. 
For pulse energies exceeding a material-dependent threshold intensity, the laser pulse releases energy into a backward-propagating Stokes wave, which is frequency-shifted by a few GHz relative to the incoming wave \cite{Chiao1964, Armandillo1983, Agrawal2013}. SBS thus represents a potential loss channel for an nMPC. The SBS frequency shift is a material-dependent property. For FS this shift is equivalent to \SI{16.3}{GHz} \cite{Bai2018}. 
Following the formalism of transient Brillouin scattering ($\tau<\SI{100}{ns}$) in fibers \cite{Keaton2014}, it can be shown that the effective intensity for the nMPC example reported above is two orders of magnitude lower than the threshold intensity required for SBS, i,e., $I^{\text{eff}}_{\text{nMPC}} / I^\text{th}_{\text{SBS}} = 0.013$ (see Figure \ref{fig:brillouin_estimate}). For the same intensity, the SBS threshold is instead surpassed when the input pulse duration is increased to $\tau \approx \SI{5}{ns}$. 
In this case, SBS can be suppressed by reducing the peak power at the expense of the single-pass spectral broadening. 

\begin{figure}[ht!]
    \centering
    \captionsetup{width=\linewidth}
    \includegraphics[width=\textwidth]{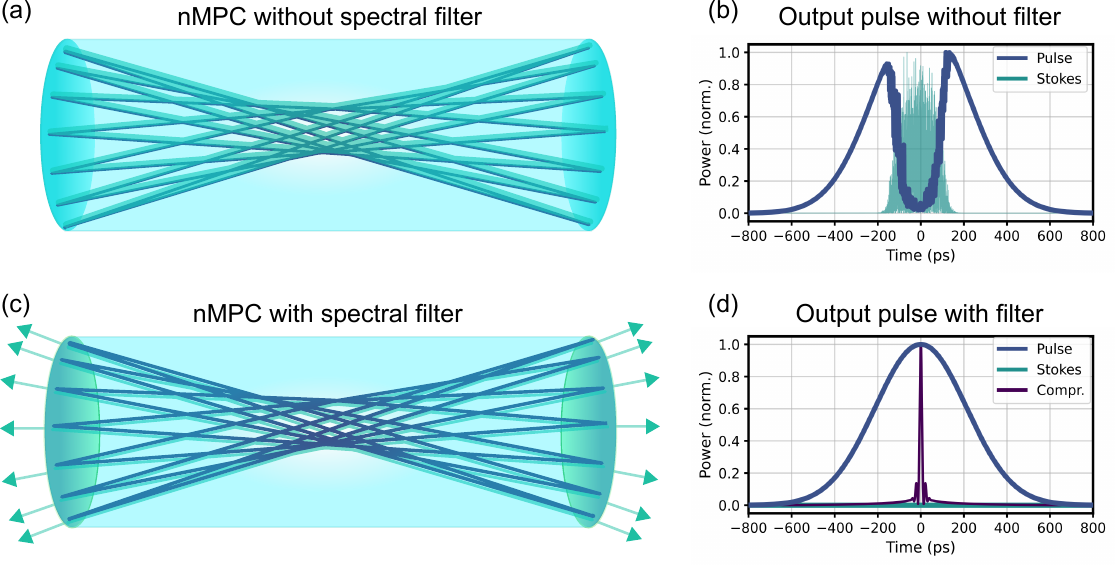}
    \caption{\textbf{Impact and suppression of SRS in nMPC post-compression.} (a,c) Illustration of the nMPC with (c) and without (a) spectral SRS filter. The transmitted Stokes components are pictorially indicated with arrows. (b,d) Corresponding simulated temporal pulse shapes following spectral broadening in the nMPC. Without filter, the pulse transfers energy to the Stokes field, leading to a distorted temporal pulse shape (b). This effect can be suppressed (d) using a suitable MPC mirror design, acting as spectral filter suppressing the Stokes pulse.}
    \label{fig:raman_filter}
\end{figure}

In addition to SBS, Stimulated Raman Scattering (SRS) can severely limit spectral broadening. 
SRS is an inelastic, stimulated scattering process in which photons are downshifted in energy by an amount determined by the vibrational modes of the medium \cite{Agrawal2013}.
In fused silica, the Stokes wave is frequency-down-shifted by about \SI{13.6}{THz} \cite{Stolen1989}.
Similar to SBS, the magnitude of SRS becomes important after exceeding a certain intensity threshold.
Due to the large frequency shift, dispersion between the original pump pulse and the Stokes pulse can influence the dynamics of SRS \cite{Agrawal2013}.
SRS has been the major limiting factor of spectral broadening in fibers \cite{Nakashima1987}.
Although large broadening factors up to 80 have been demonstrated in fibers \cite{Johnson1984, Kafka1984}, SRS limits the possible pulse energies to far below \SI{1}{\micro J}.
While in our proof-of-principle experiment we did not observe any significant SRS onset, in the case of the upscaled simulation reported in Fig. \ref{fig:outlook}, SRS becomes prominent (see eq. (\ref{eq:raman_threshold})).
However, within the nMPC scheme the SRS contribution  can be suppressed by using suitable MPC mirror coatings, acting as spectral filters for the Stokes pulse at each reflection.

Using fused silica as nonlinear material, the Stokes pulse of a laser pulse centered at \SI{1064}{nm} appears at around \SI{1112}{nm} and thus far outside the spectral range covered by the SPM-broadened spectrum, making spectral filtering possible. 
Figure \ref{fig:raman_filter} shows two simulated spectral broadening scenarios in an nMPC using \SI{500}{ps}, \SI{1}{mJ} pulses.
Attenuating the Stokes field by only 30\% per mirror reflection completely suppresses SRS and results in clean, compressible pulses (Fig. \ref{fig:raman_filter}(c, d)).

Another limiting nonlinear effect is quasi-phase-matched degenerate four-wave-mixing (QPM-FWM), which can occur in MPCs with large broadening factors \cite{Hanna2020}.
Similar to SRS, this effect can severely limit spectral broadening due to the generation of spectral side-bands, introducing strong modulations on the pulse in the temporal domain.
We analyze this effect numerically and estimate its impact for nMPC post-compression (further details in Methods section), identifying a simple mitigation strategy. 
The location of the spectral side-bands depends on the dispersion properties of the nMPC, including the nonlinear medium and the mirrors, as well as on the MPC configuration. For the considered example case with the laser pulse centered at \SI{1064}{nm}, the first QPM-FWM peaks appear at \SI{1047}{nm} and \SI{1082}{nm}.
Similar to SRS, suitable optical coatings can be used to filter out both QPM-FWM induced spectral side-bands at each reflection and fully suppress their effect (see Figure \ref{fig:fwm_nofilter}).

In the presented experiment, the moderate broadening factor of 20 resulted in the absence of the majority of the aforementioned detrimental effects, including SRS and QPM-FWM. However, our simulations reveal the importance of SRS and QPM-FWM suppression via spectral filtering for large spectral broadening factors (Figures \ref{fig:raman_SPM}, \ref{fig:fwm_nofilter}), indicating the importance of suitable mitigation strategies. We note that the (2+1)D simulation results presented in Figures \ref{fig:outlook} and \ref{fig:parabola_pulse} do not include the effect of SRS and QFM-FWM for computational reasons. However, we also performed (1+1)D simulations with the same parameters, while also considering SRS and QFM-FWM (see Figure \ref{fig:fwm_nofilter}). In particular, we show that the use of a suitable optical coating effectively filters out and suppresses the contribution of SRS and QFM-FWM even for large compression factors.\\

\section*{Conclusion and Outlook}
We introduced a novel approach to compress Q-switched laser pulses using an nMPC, which uses a long glass rod as the nonlinear medium, as well as spectral filtering concepts to suppress detrimental nonlinear effects. Our approach enables efficient spectral broadening of low peak-power sources and extremely high compression factors, while also overcoming key limitations observed in fiber-based spectral broadening, such as SRS. Our proof-of-concept experiment demonstrates the compression of a \SI{0.5}{ns} long pulse to a duration of \SI{24}{ps}, with numerically validated scalability down to \SI{3}{ps}. When combined with a second post-compression stage, the nMPC provides a bridge between nanosecond Q-switched lasers and femtosecond pulses. Our results open perspectives towards low-cost, high-peak-power ultrafast sources. Ultimately, the nMPC technology is promising to support a range of advanced applications, including the generation of high-flux extreme-ultraviolet (EUV) light for material characterisation and ultrafast laser spectroscopy research, high-precision laser-driven material machining for industrial manufacturing, and damage-free tissue treatment in dermatology. 

\newpage
\vspace{5mm}
\section*{Methods}

\FloatBarrier
\subsection*{Characterising the beam quality}
\label{supl:beam_quality}

\begin{figure}[ht!]
    \centering
    \captionsetup{width=\linewidth}
    \includegraphics[width=\textwidth]{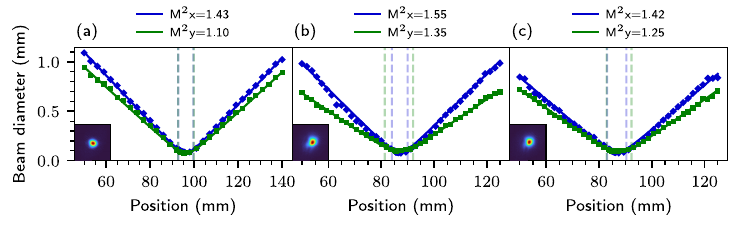}
    \caption{\textbf{Spatial beam quality characterization.} Beam quality measurement of MPC input (a), output (b) and output without Etalon at \SI{1.35}{mJ} input energy (c). The panels include the corresponding beam profiles at the focus position (insets).}
    \label{fig:beam_results}
\end{figure}

Maintaining the beam quality is an important characteristic of an MPC. In our experiment, the beam quality is measured before and after the nMPC, showing a slight degradation of the mean M$^2$ from 1.31 to 1.45 (Fig. \ref{fig:beam_results}). This is potentially caused by the beam passing through the surface of the FS bulk with an angle at each pass. This problem could potentially be mitigated by the proposed all-bulk MPC, which would eliminate the need for surface passes. We observed an improving output beam quality with increasing power.

\FloatBarrier
\subsection*{Influence of additional longitudinal laser modes in post-compression schemes}
\label{supl:LM}

Unlike mode-locked lasers, Q-switched lasers typically emit non-phase-locked secondary longitudinal modes (LM). These modes can distort the SPM process as secondary modes cause distortions of the temporal pulse envelope. This can be critical even if only a very small fraction of the energy is emitted in weak additional modes. For the passively Q-switched laser used in our experiment, the manufacturer specifies the single LM operation as having an energy part of less than \SI{0.1}{}\% in different LMs. In our simulation, we observed good agreement with the experimental results for a single second mode that contained \SI{0.01}{}\% of the pulse energy.

\begin{figure}[ht!]
    \centering
    \captionsetup{width=\linewidth}
    \includegraphics[width=.95\textwidth]{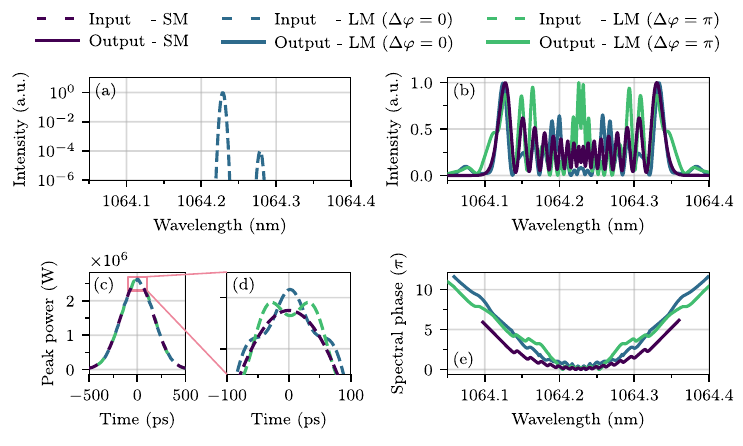}
    \caption{\textbf{Simulation results visualizing the influence of a secondary longitudinal mode on the compressibility of the MPC output.} (a) Input spectrum with a secondary longitudinal mode (LM) with \SI{0.01}{}\% of the energy (b) Spectral broadening performance for single longitudinal mode operation (see Methods section) and for secondary LM operation. The broadening is shown for two different relative phases $\Delta\varphi$ between the LMs. (c,d) Temporal Intensity profile, with oscillations caused by interference with the second LM. (e) From SPM generated spectral phase, which has to be compensated for compression.}
    \label{fig:2nd_LM}
\end{figure}

Figure \ref{fig:2nd_LM}(a) shows an example of a single secondary LM separated by \SI{50}{pm}. Although the interference modulates the temporal pulse envelope by only a few percent (Fig. \ref{fig:2nd_LM}(c, d)), our simulations reveal that the broadened spectra are distorted with a strong dependence on the respective inter-mode phase (Fig. \ref{fig:2nd_LM}(b)), compared to the single-mode case. 
Figure \ref{fig:2nd_LM}(d) displays the spectral phase of the three cases, where only the single-mode operation indicates a spectral phase that fits a parabola closely, thus enabling compression by GDD compensation.

\begin{figure}[ht!]
    \centering
    \captionsetup{width=\linewidth}
    \includegraphics[width=.95\textwidth]{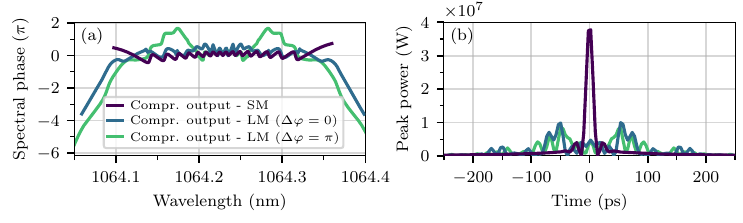}
    \caption{\textbf{Influence of secondary LM's on compressibility of the MPC output.} (a) Remaining spectral phase after GDD compensation. (b) Corresponding compressed output pulses.}
    \label{fig:2nd_LM_comp}
\end{figure}

Figure \ref{fig:2nd_LM_comp}(a) shows the spectral phase after GDD compensation. It is evident that the outer spectral components, in particular, hinder successful pulse compression. The corresponding compressed pulses are displayed in Figure \ref{fig:2nd_LM_comp}(b), where temporal break-up is observed in cases where a weak second mode is present. This demonstrates that, to achieve high compression factors and a well-defined output pulse, the system must operate in a single mode. In this work, single-mode operation is ensured by using a Fabry-Pérot etalon. In our experiment, the effect discussed here was observed, and it prevented compression close to the Fourier transform limit (FTL) in the absence of the etalon.

\FloatBarrier
\subsection*{Estimation of Stimulated Brillouin Scattering (SBS)}
\label{supl:SBS}

While SBS can be used as the main nonlinear process for pulse compression of long pulses - typically \SI{10}{ns} or longer down to below $\SI{1}{ns}$ \cite{Gen300psSBS, HighESBS} - it is detrimental for spectral broadening through SPM. For pulses that are longer than the phonon lifetime of the material $\tau \gg T_B$ (for fused silica $T_B\simeq\SI{5}{ns}$), the SBS threshold peak intensity $I_\text{th}$ can be calculated using the steady-state SBS gain coefficient $g_B$ and the threshold parameter $\Theta$ \cite{Chen2023}. This material-dependent quantities and the interaction length $L$ define the single-pass peak intensity threshold as \cite{Keaton2014}:

\begin{equation}\label{eq:brillouin_threshold}
    I_\text{th}=\frac{\Theta}{ g_B \cdot L } 
        \approx \frac{\Theta}{ g_B \cdot z_R }, 
    \quad\tau \gg 2z_R / v_\mathrm{g}, 
\end{equation}

\noindent
Here $L\approx z_{\mathrm{R}}$ denotes the interaction length, $z_{\mathrm{R}}$ the Rayleigh length and $v_{\mathrm{g}}$ the group velocity of the pulse. 
For the transient Brillouin scattering regime, i.e., when the pulse duration is shorter than the phonon lifetime, the temporal interaction decreases, thus increasing the threshold intensity. In this regime, SBS can be modeled using numerical methods that have been developed for optical fibers. Here, the threshold peak intensity scales with pulse duration as \cite{Keaton2014}:

\begin{equation}\label{eq:brillouin_threshold_pulsed}
     I_\text{th}=\frac{2 \Theta}{g_B v_{\mathrm{g}} \tau} \left(\frac{T_B \Theta}{\tau} + 1\right),
     \quad\tau \le 2L/v_{\mathrm{g}}.
\end{equation}

\noindent
In order to quantify this intensity threshold for the nMPC, we define an effective intensity $I_{\text{eff}}$ of a single pass through the nMPC, that would produce the same nonlinearity, i.e., the same $B$-integral, as propagation in a fiber with the corresponding constant mode size:

\begin{equation}
    B_\text{pass} = \frac{2\pi}{\lambda} n_2 L_\mathrm{MPC} I_\text{eff}
    \label{eq:B_const_I}
\end{equation}

\noindent
Considering the basic geometry of an MPC filled with a nonlinear medium, the $B$-integral can be written as \cite{MPCReview}:

\begin{equation}
    B_\text{pass} = 4\pi^2 \frac{n_2 P}{\lambda^2} \frac{k}{N}
    \label{eq:B_MPC}
\end{equation}

\noindent
Using equations \ref{eq:B_const_I} and \ref{eq:B_MPC}, we obtain the effective intensity:

\begin{equation}
     I_{\text{eff}} = \frac{\pi P \, n}{\lambda \, L_{\mathrm{MPC}}} \frac{k}{N}
     \label{eq:Ieff_sbs}
\end{equation}

\noindent
Here, $P$ denotes the pulse peak power and $n$ the refractive index. The MPC configuration is defined by the cell length $L_{\mathrm{MPC}}$, the number of round trips $N$, and by the re-entrance condition defined by the integer $k$. For a given set of material parameters, the SBS threshold intensity can be calculated in terms of interaction length (eq. \ref{eq:brillouin_threshold}) and pulse duration (eq. \ref{eq:brillouin_threshold_pulsed}). This threshold can then be compared to the effective Intensity in the MPC using equation \ref{eq:Ieff_sbs}.

\begin{figure}[ht!]
    \centering
    \captionsetup{width=\linewidth}
    \includegraphics[width=.95\textwidth]{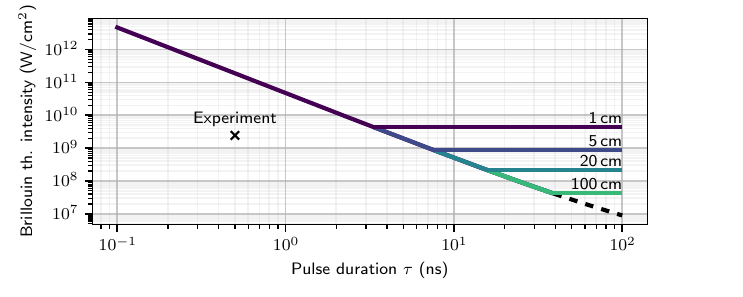}
    \caption{\textbf{SBS threshold in an nMPC}. The dependence of the Brillouin threshold intensity is plotted as a function of the input pulse duration (eq. \ref{eq:brillouin_threshold_pulsed}). For long pulses, the threshold is dominated by the interaction length (eq. \ref{eq:brillouin_threshold}). Our experimental parameters are far below the SBS threshold (eq. \ref{eq:Ieff_sbs}). The following parameters were used: $g_B = \SI{5}{cm/GW}$, $T_B = \SI{5}{ns}$, $n_\text{FS} = 1.45$, $\Theta = 22$ and $n_2 = \SI{2.19e-16}{cm^2/W}$ \cite{Agrawal2013, Boyd2020, Keaton2014, Kabacinski:19}.}
    \label{fig:brillouin_estimate}
\end{figure}

Figure \ref{fig:brillouin_estimate} shows the impact of SBS in a fused silica nMPC for input pulse durations from \SI{0.1}{ns} to \SI{100}{ns}.
For parameters that lay under the curves in Figure \ref{fig:brillouin_estimate}, the contribution of SBS is negligible, as in the case of our experimental demonstration. For the same MPC configuration as used in the experiment, a sufficient single-pass B-integral of at least $\geq 0.1\pi$ could be achieved with an input laser pulse duration of up to $\SI{6}{ns}$. For longer pulse durations, a large number of passes would be needed to compensate for the low single-pass B-integral. Alternatively, the SBS component needs to be attenuated after each pass.

\FloatBarrier
\subsection*{Simulation of Stimulated Raman scattering (SRS)}
\label{supl:SRS}

The threshold intensity above which SRS becomes significant depends on the pulse duration and material-dependent parameters. It can be heuristically defined as \cite{Agrawal2013}:

\begin{equation}
    I_{\mathrm{th}}^{\mathrm{SRS}}  = \ \frac{32}{L_{\mathrm{W}} \, g_{\mathrm{R}} } \ .
    \label{eq:raman_threshold}
\end{equation}

\begin{figure}[hb!]
    \centering
    \captionsetup{width=\linewidth}
    \includegraphics[width=.95\textwidth]{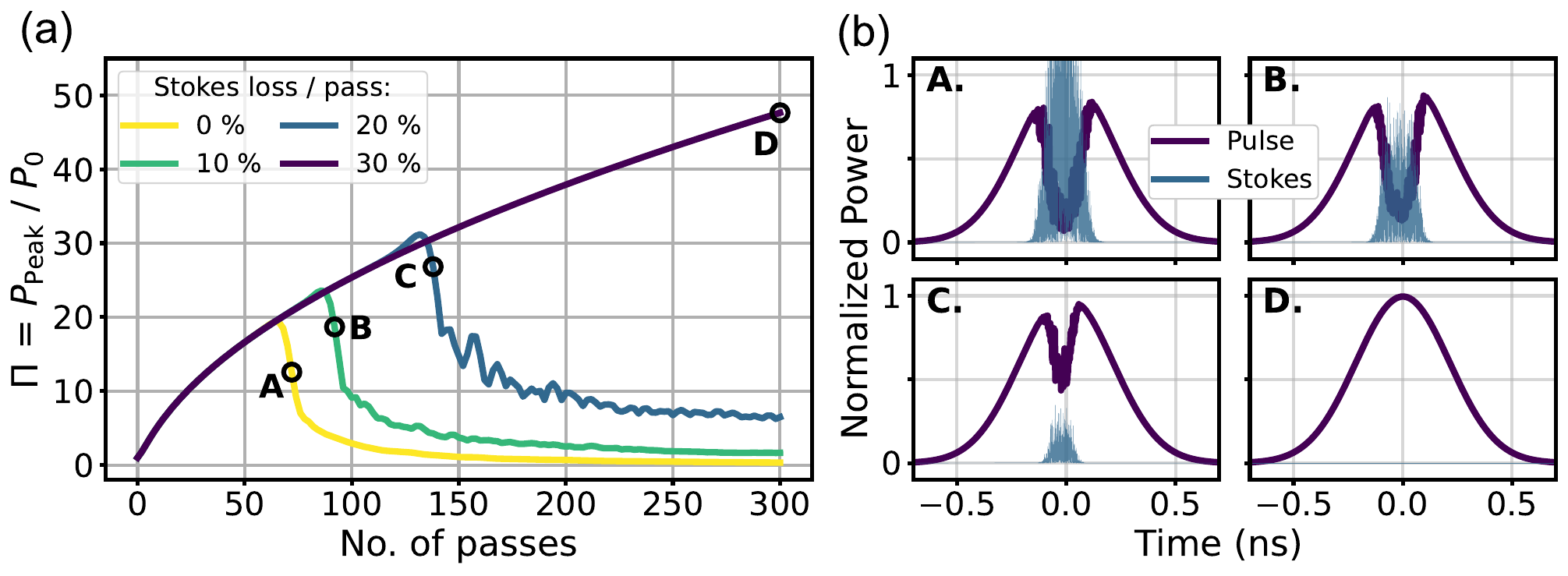}
    \caption{\textbf{Impact of SRS in an nMPC for different Stokes losses}. (a) Simulated peak power enhancement $\Pi = P_{\mathrm{peak}}\,/\,P_0$ after spectral broadening of \SI{1}{mJ}, \SI{0.5}{ns} pulses in an all-bulk fused silica nMPC with $N=150$, $k=91$, $R=\SI{0.1}{m}$. 
    $P_{\mathrm{peak}}$ is calculated by applying the optimized GDD to the output pulses. $P_0$ denotes the input peak power. The figure displays $\Pi$ for different values of loss applied per pass to the stokes pulse. (b) Uncompressed pulses and Stokes fields for different Stokes losses corresponding to the data displayed in (a). The displayed data correspond to different number of passes in the nMPC, as indicated with circles in (a).}
    \label{fig:raman_SPM}
\end{figure}

\noindent
Here, $g_{\mathrm{R}}$ is the Raman gain parameter, which is approximately $g_{\mathrm{R}} = \SI{1e-13}{m/W}$ for $\SI{1}{\micro m}$ wavelength in fused silica \cite{Agrawal2013}.
$L_{\mathrm{W}} = \tau/|v_{\mathrm{gp}}^{-1} - v_{\mathrm{gs}}^{-1}|$ is the so-called walk-off length, after which the pump and Stokes pulses temporally walk off, thus limiting further Raman gain with $v_{\mathrm{gp}}$, $v_{\mathrm{gs}}$ defining the group velocities of the pump and the generated Stokes pulse, respectively. 
In the case of nanosecond pulses, $L_{\mathrm{W}}$ is long and $I_{\mathrm{th}}^{\mathrm{SRS}}$ small, thus defining a severe constraint for the achievable SPM spectral broadening.
However, as shown in the main manuscript and in Fig. \ref{fig:raman_SPM}, suitably nMPC mirrors can act as a spectral filter introducing a loss for the Stokes wave. A moderate SRS attenuation of 30\% per pass is already sufficient to fully suppress the overall SRS contribution. Considering the THz-level frequency shift of the SRS Stoke wave, the production of such optical coatings is fully feasible\cite{Paschotta_2005_dielectric_mirrors}.

Below, we describe the numerical model used for the SRS simulation.
For computational cost reasons, we use a (1+1)D model that takes temporal effects into account, while the spatial dimension is modeled using the ray-transfer matrix formalism.
For the nonlinear propagation including SRS, we follow the same approach as described in ref. \cite{Headley1996} (eqs. (10) and (11)), defining the two coupled equations for pump and Stokes fields. We modify the equations slightly, omitting the last term in both equations, which describes spontaneous Raman scattering. Further, we include the full linear dispersion. The two coupled equations read as:

\begin{align*}
\frac{\partial u_p}{\partial z} + i D u_p
&= i \gamma_p (1 - f_R) u_p \left( |u_p|^2 + 2|u_s|^2 \right) 
+ \\
&\quad + \ i \gamma_p f_R u_p  \int_{-\infty}^\infty h_R(t - t') \left[ |u_p(t')|^2 + |u_s(t')|^2 \right] dt' \\
&\quad + \ i \gamma_p f_R u_s  \int_{-\infty}^\infty h_R(t - t') u_p(t') u_s^*(t') \exp[i\Omega_R (t - t')] dt'
\end{align*}

\begin{align*}
\frac{\partial u_s}{\partial z} +  i D u_s
&= i \gamma_s (1 - f_R) u_s \left( |u_s|^2 + 2|u_p|^2 \right) 
+ \\
&\quad + \ i \gamma_s f_R u_s  \int_{-\infty}^\infty h_R(t - t') \left[ |u_s(t')|^2 + |u_p(t')|^2 \right] dt' \\
&\quad + \ i \gamma_s f_R u_p \int_{-\infty}^\infty h_R(t - t') u_s(t') u_p^*(t') 
\exp[-i\Omega_R (t - t')] dt' \ ,
\end{align*}

\noindent
where $u_i$, $i=p,s$ (pump and Stokes, respectively) are the fields in the temporal domain, normalized such that the power $P_i(t) = |u_i|^2 $. 
The dispersion operator $D$ includes all dispersion orders, incorporated by solving the left-hand side of the equations in the frequency domain and using refractive index data of fused silica \cite{Malitson1965}. 
Further, $\gamma_i$ is the nonlinear coupling parameter which is proportional to the nonlinear refractive index $n_2$, defined in ref. \cite{Headley1996}. 
We use $n_2=\SI{2.19e-20}{m^2/W}$ \cite{Kabacinski2019} for the nonlinear refractive index of fused silica. 
Finally, $f_{\mathrm{R}} = 0.18$ is the Raman fraction of fused silica \cite{Headley1996}, and $h_{\mathrm{R}}(t)$ is the delayed response function of the SRS interaction in the temporal domain. 

\begin{figure}[ht!]
    \captionsetup{width=\linewidth}
    \centering
    \includegraphics[width=0.7\textwidth]{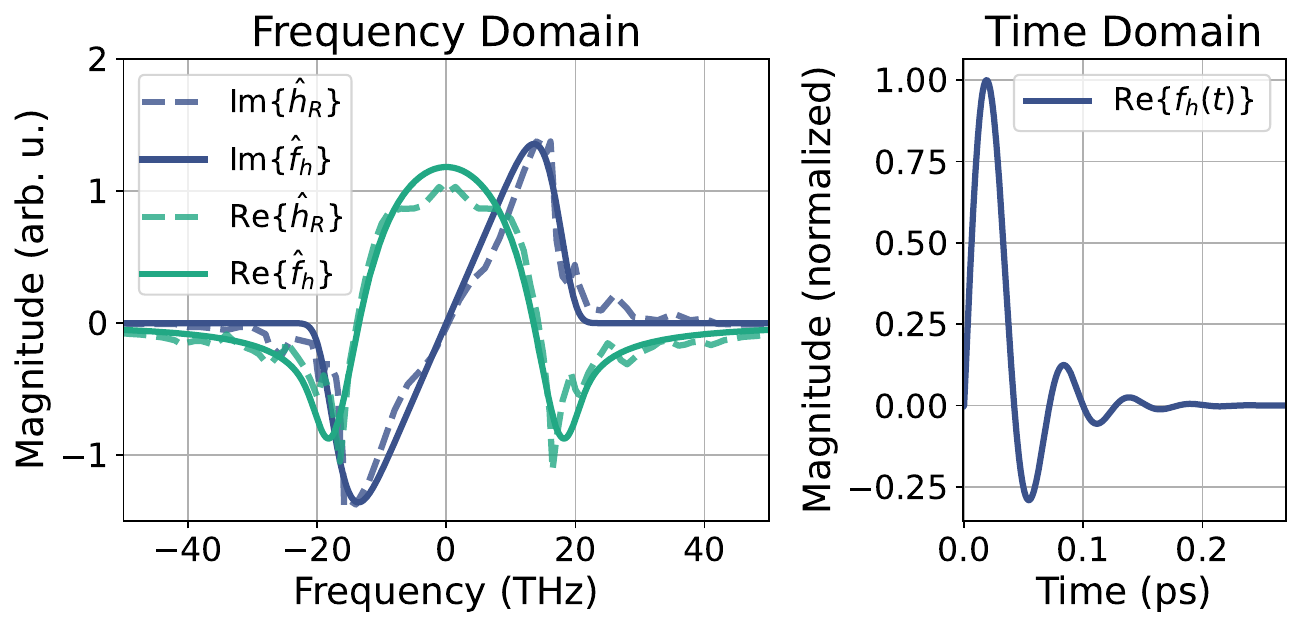}
    \caption{\textbf{Raman gain of fused silica}. The data for the delayed response function of the SRS interaction in frequency domain $\hat{h}_R(\omega)$ are taken from ref. \cite{Headley1996}. The fitted functions $\hat{f}_h(\omega)$ in the frequency domain and $f_h(t)$ in the time domain are used in the simulations.}
    \label{fig:raman_gain}
\end{figure}

To determine $h_{\mathrm{R}}(t)$, we use the data for the response function in the frequency domain $\hat{h}_R(\omega)$ provided in ref. \cite{Headley1996, Stolen1989} and fit a simplified model $f_h(\omega) = a\cdot\omega\cdot\exp{[-(b\cdot\omega)^8]}$ with the fitting parameters  $a,\ b$ to the imaginary part.
We calculate the real part using the Kramers-Kronig relationship, while ensuring normalization
$\int_{-\infty}^{{-\infty}}{f_h(t)\,dt} = 1$ \cite{Headley1996}. 
Figure \ref{fig:raman_gain} shows the response functions in the temporal and frequency domains.

The Stokes field is initialized with a random noise following a Gaussian distribution using the one-photon-per-mode method \cite{Agrawal2013}.
We solve the model using a coupled Runge-Kutta-4 split-step algorithm on a $2^{18} = 262144$ temporal/spectral grid with a temporal window of $[-1.27, 1.27] \ \mathrm{ns}$ and a step size $dz=0.00246 \ \mathrm{m}$.

\FloatBarrier
\subsection*{Quasi-Phase-Matched Four-Wave-Mixing (QPM-FWM) simulations}
\label{supl:FWM}

\begin{figure}[ht]
    \captionsetup{width=\linewidth}
    \centering
    \includegraphics[clip,width=0.9\textwidth]{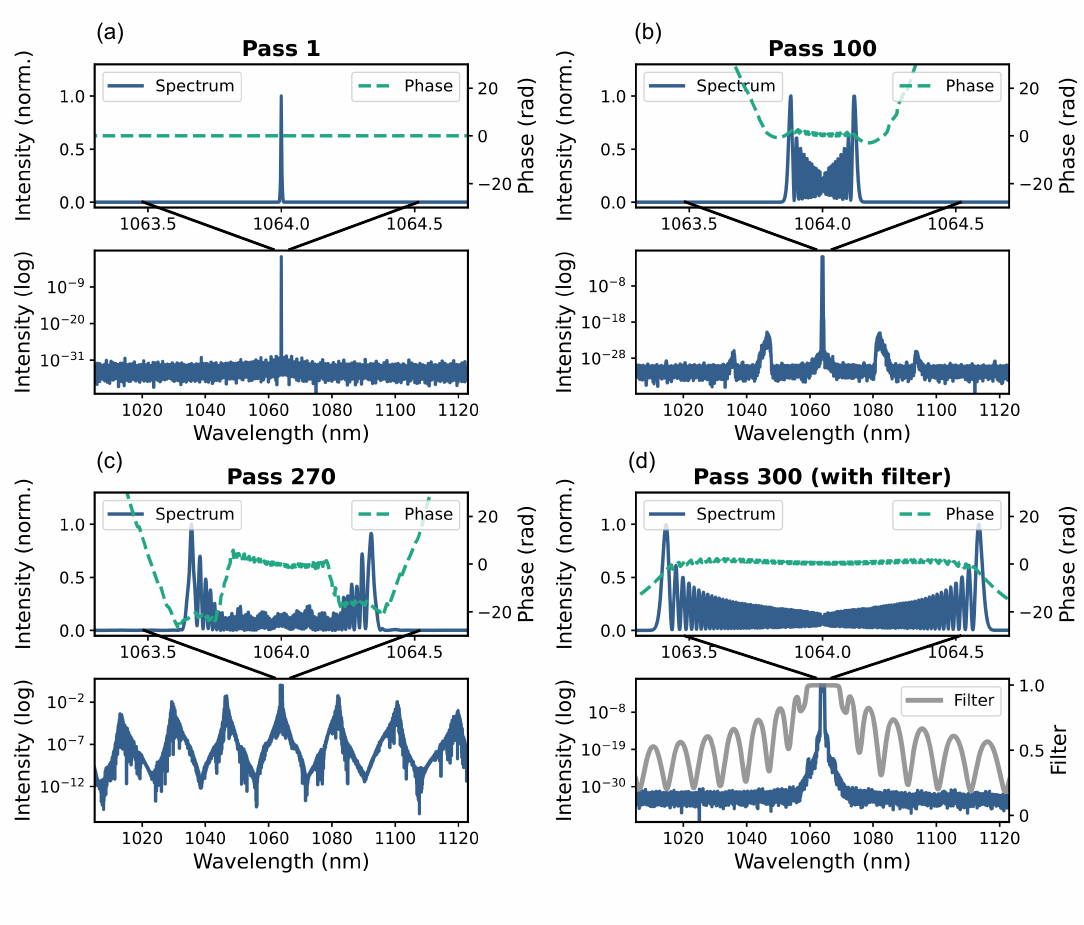}
    \caption{\textbf{Impact of QPM-FWM in an nMPC}. The QPM-FWM contribution is simulated after (a) 1, (b) 100, and (c) 270 passes without using a spectral filter. (d) Plot of the broadened spectrum applying a spectral filter after each pass.
    In each of the Figures (a)-(d), the spectrum is plotted in linear scale (upper plot) and logarithmic scale (lower plot, zoomed view of upper plot).The spectral phase is plotted after subtracting the GDD, optimized for the highest peak power. The gray line in (d), lower panel visualizes the spectral filter.}
    \label{fig:fwm_nofilter}
\end{figure}

Quasi-Phase-Matched Four-Wave-Mixing (QPM-FWM) can occur at large spectral broadening factors in MPCs, as discussed in ref. \cite{Hanna2020}. 
Following ref. \cite{Hanna2020}, we performed (1+1)D simulations to quantify the possible contribution of QPM-FWM in a nMPC. A \SI{0.5}{ns}, \SI{1}{mJ} pulse is propagated in a full-bulk fused silica nMPC, with $R=\SI{10}{cm}$, $N=150$, $k=91$, resulting in $2N = 300$ passes.
Figure \ref{fig:fwm_nofilter}(a)-(c) shows the impact of QPM-FWM on the broadened spectrum after 1, 100, and 270 passes through the nMPC. Spectral side bands appear at specific wavelengths, which are determined by the dispersion landscape of the nMPC. 
More specifically, QPM-FWM occurs at wavelengths for which the phase-matching condition $\Delta\phi = (2k_1 - k_2 - k_3) L_{\mathrm{MPC}} = n2\pi$ ($n$ integer) applies after one pass, where $k(\omega_i) \equiv k_i$ is the wave-number, $i=1$ denotes the fundamental wavelength and $i=2,3$ the FWM components \cite{Hanna2020}.

Figure \ref{fig:fwm_nofilter}(c) clearly shows a large energy transfer to the FWM spectral components and a distortion of the main spectrum.
Furthermore, the phase exhibits discontinuities which cannot be compensated via simple removal of group-delay-dispersion (GDD), resulting in an incompressible pulse in the time domain.
To suppress both SRS and QPM-FWM, we apply a spectral filter at each reflection at the nMPC mirror in the simulation.
Figure \ref{fig:fwm_nofilter}(d) shows that the contributions of QPM-FWM and SRS are fully suppressed by the spectral filter, resulting in a clean and compressible SPM-broadened spectrum.
The spectral filter consists of two dielectric multi-layer mirror coatings, one with a central wavelength offset to the longer and one to the shorter side of the spectrum.
The effective filter per round-trip in the nMPC is shown in Fig. \ref{fig:fwm_nofilter}(d).
After GDD removal, the phase is smooth without discontinuities. 

\FloatBarrier
\small 
\subsection*{Acknowledgements} 
We acknowledge financial support from the European Research Council under the ERC SoftMeter no. 101076500. Views and opinions expressed are however those of the author(s) only and do not necessarily reflect those of the European Union or the European Research Council Executive Agency. Neither the European Union nor the granting authority can be held responsible for them. 

MK, UM, and AT acknowledge support by the Cluster of Excellence PhoenixD (Photonics, Optics, and Engineering-Innovation Across Disciplines), DFG EXC 2181. We also acknowledge funding from the Cluster of Excellence ‘CUI: Advanced Imaging of Matter’ of the Deutsche Forschungsgemeinschaft (DFG) – EXC 2056 – project ID 390715994. PBi acknowledges support from the European Innovation Council contract EIC open “NanoXCAN” (Grant No. 101047223). AT acknowledges support from the Helmholtz Association under the Helmholtz Young Investigator Group VH-NG-1603. CH acknowledges support from the Deutsche Forschungsgemeinschaft (DFG, German Research Foundation) – Project 545612524. AS is supported by the Helmholtz-Lund International Graduate School (HELIOS), Helmholtz Project Number HIRS-0018.

We further acknowledge Leibniz University Hannover (Hannover, Germany), DESY (Hamburg, Germany) and the Helmholtz-Institute Jena (Jena, Germany) members of the Helmholtz Association HGF, for support.

\subsection*{Author contributions}
CH and AT conceived the initial idea and supervised the project. PBi, AS, MS, NJ, TL, JKS, SF, PM performed the experiment. PBi and AS carried out the data analysis, as well as the (1+1)D and the (2+1)D simulations supporting the concept and the experiment. PBi, AS, MS, CH, and AT drafted to writing the manuscript. All authors discussed the results and contributed to the manuscript.

\subsection*{Data availability}
The data that support the findings of this study are available from the corresponding author on request.

\subsection*{Conflict of interest}
The authors declare no competing interests.

\FloatBarrier
\bibliography{bibliography}

\begin{thebibliography}{10}
\expandafter\ifx\csname url\endcsname\relax
  \def\url#1{\texttt{#1}}\fi
\expandafter\ifx\csname urlprefix\endcsname\relax\def\urlprefix{URL }\fi
\providecommand{\bibinfo}[2]{#2}
\providecommand{\eprint}[2][]{\url{#2}}

\bibitem{RefrEyeSurg}
\bibinfo{author}{Linz, N.}, \bibinfo{author}{Freidank, S.},
  \bibinfo{author}{Liang, X.-X.} \& \bibinfo{author}{Vogel, A.}
\newblock \emph{\bibinfo{title}{{Laser Micro- and Nanostructuring for
  Refractive Eye Surgery}}}, \bibinfo{pages}{1217--1245}
  (\bibinfo{publisher}{Springer International Publishing},
  \bibinfo{address}{Cham}, \bibinfo{year}{2023}).
\newblock \urlprefix\url{https://doi.org/10.1007/978-3-031-14752-4_33}.

\bibitem{Medicine}
\bibinfo{author}{Skorczakowski, M.} \emph{et~al.}
\newblock \bibinfo{title}{Mid-infrared {Q-switched} {Er:YAG} laser for medical
  applications}.
\newblock \emph{\bibinfo{journal}{Laser Physics Letters}}
  \textbf{\bibinfo{volume}{7}}, \bibinfo{pages}{498} (\bibinfo{year}{2010}).
\newblock \urlprefix\url{https://dx.doi.org/10.1002/lapl.201010019}.

\bibitem{GasSensing}
\bibinfo{author}{Hodgkinson, J.} \& \bibinfo{author}{Tatam, R.~P.}
\newblock \bibinfo{title}{Optical gas sensing: a review}.
\newblock \emph{\bibinfo{journal}{Measurement Science and Technology}}
  \textbf{\bibinfo{volume}{24}}, \bibinfo{pages}{012004}
  (\bibinfo{year}{2012}).
\newblock \urlprefix\url{https://dx.doi.org/10.1088/0957-0233/24/1/012004}.

\bibitem{MaterProces}
\bibinfo{author}{Sugioka, K.} \& \bibinfo{author}{Cheng, Y.}
\newblock \bibinfo{title}{Ultrafast lasers – reliable tool for advanced
  material pocessing}.
\newblock \emph{\bibinfo{journal}{Light: Science \& Applications}}
  \textbf{\bibinfo{volume}{3}} (\bibinfo{year}{2014}).
\newblock \urlprefix\url{https://doi.org/10.1038/lsa.2014.30}.

\bibitem{NeurImage}
\bibinfo{author}{Fan, S.}, \bibinfo{author}{Wang, S.}, \bibinfo{author}{Yang,
  C.}, \bibinfo{author}{Wise, F.} \& \bibinfo{author}{Kong, L.}
\newblock \bibinfo{title}{Advances of mode-locking fiber lasers in neural
  imaging}.
\newblock \emph{\bibinfo{journal}{Advanced Optical Materials}}
  \textbf{\bibinfo{volume}{11}}, \bibinfo{pages}{2202945}
  (\bibinfo{year}{2023}).
\newblock
  \urlprefix\url{https://advanced.onlinelibrary.wiley.com/doi/abs/10.1002/adom.202202945}.
\newblock
  \eprint{https://advanced.onlinelibrary.wiley.com/doi/pdf/10.1002/adom.202202945}.

\bibitem{ChargeMigr}
\bibinfo{author}{Cederbaum, L.} \& \bibinfo{author}{Zobeley, J.}
\newblock \bibinfo{title}{Ultrafast charge migration by electron correlation}.
\newblock \emph{\bibinfo{journal}{Chemical Physics Letters}}
  \textbf{\bibinfo{volume}{307}}, \bibinfo{pages}{205--210}
  (\bibinfo{year}{1999}).
\newblock
  \urlprefix\url{https://www.sciencedirect.com/science/article/pii/S0009261499005084}.

\bibitem{Superconductivity}
\bibinfo{author}{Nicoletti, D.} \emph{et~al.}
\newblock \bibinfo{title}{{Optically induced superconductivity in striped
  ${\mathrm{La}}_{2\ensuremath{-}x}{\mathrm{Ba}}_{x}{\mathrm{CuO}}_{4}$ by
  polarization-selective excitation in the near infrared}}.
\newblock \emph{\bibinfo{journal}{Phys. Rev. B}} \textbf{\bibinfo{volume}{90}},
  \bibinfo{pages}{100503} (\bibinfo{year}{2014}).
\newblock \urlprefix\url{https://link.aps.org/doi/10.1103/PhysRevB.90.100503}.

\bibitem{Thorium229(1)}
\bibinfo{author}{Tiedau, J.} \emph{et~al.}
\newblock \bibinfo{title}{{Laser Excitation of the Th-229 Nucleus}}.
\newblock \emph{\bibinfo{journal}{Phys. Rev. Lett.}}
  \textbf{\bibinfo{volume}{132}}, \bibinfo{pages}{182501}
  (\bibinfo{year}{2024}).
\newblock
  \urlprefix\url{https://link.aps.org/doi/10.1103/PhysRevLett.132.182501}.

\bibitem{Thorium229(2)}
\bibinfo{author}{Elwell, R.} \emph{et~al.}
\newblock \bibinfo{title}{{Laser Excitation of the $^{229}\mathrm{Th}$ Nuclear
  Isomeric Transition in a Solid-State Host}}.
\newblock \emph{\bibinfo{journal}{Phys. Rev. Lett.}}
  \textbf{\bibinfo{volume}{133}}, \bibinfo{pages}{013201}
  (\bibinfo{year}{2024}).
\newblock
  \urlprefix\url{https://link.aps.org/doi/10.1103/PhysRevLett.133.013201}.

\bibitem{QSwitching}
\bibinfo{author}{Wang, Y.} \& \bibinfo{author}{Xu, C.-Q.}
\newblock \bibinfo{title}{{Actively Q-switched fiber lasers: Switching dynamics
  and nonlinear processes}}.
\newblock \emph{\bibinfo{journal}{Progress in Quantum Electronics}}
  \textbf{\bibinfo{volume}{31}}, \bibinfo{pages}{131--216}
  (\bibinfo{year}{2007}).
\newblock
  \urlprefix\url{https://www.sciencedirect.com/science/article/pii/S0079672707000419}.

\bibitem{QSwitchedLaser}
\bibinfo{author}{Chen, X.}, \bibinfo{author}{Wang, N.}, \bibinfo{author}{He,
  C.} \& \bibinfo{author}{Lin, X.}
\newblock \bibinfo{title}{{Development of all-fiber nanosecond oscillator using
  actively Q-switched technologies and modulators}}.
\newblock \emph{\bibinfo{journal}{Optics \& Laser Technology}}
  \textbf{\bibinfo{volume}{157}}, \bibinfo{pages}{108709}
  (\bibinfo{year}{2023}).
\newblock
  \urlprefix\url{https://www.sciencedirect.com/science/article/pii/S0030399222008556}.

\bibitem{QSwitchSatAbs}
\bibinfo{author}{Al-Hiti, A.~S.}, \bibinfo{author}{Yasin, M.} \&
  \bibinfo{author}{Harun, S.~W.}
\newblock \bibinfo{title}{{Nanosecond Q-switched laser with PEDOT: PSS
  saturable absorber}}.
\newblock \emph{\bibinfo{journal}{Appl. Opt.}} \textbf{\bibinfo{volume}{61}},
  \bibinfo{pages}{1292--1299} (\bibinfo{year}{2022}).
\newblock
  \urlprefix\url{https://opg.optica.org/ao/abstract.cfm?URI=ao-61-6-1292}.

\bibitem{PulsedLaserGener}
\bibinfo{author}{Paschotta, R.}
\newblock \emph{\bibinfo{title}{Field Guide to Laser Pulse Generation}}
  (\bibinfo{publisher}{SPIE Press}, \bibinfo{year}{2008}).

\bibitem{HighEQSwitch}
\bibinfo{author}{Zhou, Y.} \emph{et~al.}
\newblock \bibinfo{title}{{High-pulse-energy passively Q-switched
  sub-nanosecond MOPA laser system operating at kHz level}}.
\newblock \emph{\bibinfo{journal}{Opt. Express}} \textbf{\bibinfo{volume}{29}},
  \bibinfo{pages}{17201--17214} (\bibinfo{year}{2021}).
\newblock
  \urlprefix\url{https://opg.optica.org/oe/abstract.cfm?URI=oe-29-11-17201}.

\bibitem{ModeLockUltra}
\bibinfo{author}{Kues, M.} \emph{et~al.}
\newblock \bibinfo{title}{Passively mode-locked laser with an ultra-narrow
  spectral width}.
\newblock \emph{\bibinfo{journal}{Nature Photonics}}
  \textbf{\bibinfo{volume}{11}}, \bibinfo{pages}{159--162}
  (\bibinfo{year}{2017}).
\newblock \urlprefix\url{https://doi.org/10.1038/nphoton.2016.271}.

\bibitem{RevModeLock}
\bibinfo{author}{Kim, J.} \& \bibinfo{author}{Song, Y.}
\newblock \bibinfo{title}{Ultralow-noise mode-locked fiber lasers and frequency
  combs: principles, status, and applications}.
\newblock \emph{\bibinfo{journal}{Adv. Opt. Photon.}}
  \textbf{\bibinfo{volume}{8}} (\bibinfo{year}{2016}).

\bibitem{RevModeLockFiber}
\bibinfo{author}{Han, Y.} \emph{et~al.}
\newblock \bibinfo{title}{Generation, optimization, and application of
  ultrashort femtosecond pulse in mode-locked fiber lasers}.
\newblock \emph{\bibinfo{journal}{Progress in Quantum Electronics}}
  \textbf{\bibinfo{volume}{71}}, \bibinfo{pages}{100264}
  (\bibinfo{year}{2020}).
\newblock
  \urlprefix\url{https://www.sciencedirect.com/science/article/pii/S0079672720300185}.

\bibitem{PostCompressionTech}
\bibinfo{author}{Nagy, T.}, \bibinfo{author}{Simon, P.} \&
  \bibinfo{author}{Veisz, L.}
\newblock \bibinfo{title}{High-energy few-cycle pulses: post-compression
  techniques}.
\newblock \emph{\bibinfo{journal}{Advances in Physics: X}}
  \textbf{\bibinfo{volume}{6}} (\bibinfo{year}{2020}).

\bibitem{AdvancesHCF}
\bibinfo{author}{Frosz, M.~H.} \emph{et~al.}
\newblock \bibinfo{title}{Editorial: Advances and applications of hollow-core
  fibers}.
\newblock \emph{\bibinfo{journal}{IEEE Journal of Selected Topics in Quantum
  Electronics}} \textbf{\bibinfo{volume}{30}}, \bibinfo{pages}{1--1}
  (\bibinfo{year}{2024}).

\bibitem{Nisoli1996}
\bibinfo{author}{Nisoli, M.}, \bibinfo{author}{De~Silvestri, S.} \&
  \bibinfo{author}{Svelto, O.}
\newblock \bibinfo{title}{Generation of high energy 10 fs pulses by a new pulse
  compression technique}.
\newblock \emph{\bibinfo{journal}{Applied Physics Letters}}
  \textbf{\bibinfo{volume}{68}}, \bibinfo{pages}{2793--2795}
  (\bibinfo{year}{1996}).

\bibitem{PCFsub70fs}
\bibinfo{author}{Gebhardt, M.} \emph{et~al.}
\newblock \bibinfo{title}{{Nonlinear compression of an ultrashort-pulse
  thulium-based fiber laser to sub-70\si{fs} in Kagome photonic crystal
  fiber}}.
\newblock \emph{\bibinfo{journal}{Opt. Lett.}} \textbf{\bibinfo{volume}{40}},
  \bibinfo{pages}{2770--2773} (\bibinfo{year}{2015}).
\newblock
  \urlprefix\url{https://opg.optica.org/ol/abstract.cfm?URI=ol-40-12-2770}.

\bibitem{PCFsub2Cycle}
\bibinfo{author}{Heidt, A.~M.} \emph{et~al.}
\newblock \bibinfo{title}{High quality sub-two cycle pulses from compression of
  supercontinuum generated in all-normal dispersion photonic crystal fiber}.
\newblock \emph{\bibinfo{journal}{Opt. Express}} \textbf{\bibinfo{volume}{19}},
  \bibinfo{pages}{13873--13879} (\bibinfo{year}{2011}).
\newblock
  \urlprefix\url{https://opg.optica.org/oe/abstract.cfm?URI=oe-19-15-13873}.

\bibitem{PCFKagome}
\bibinfo{author}{Mak, K.~F.}, \bibinfo{author}{Travers, J.~C.},
  \bibinfo{author}{Joly, N.~Y.}, \bibinfo{author}{Abdolvand, A.} \&
  \bibinfo{author}{Russell, P. S.~J.}
\newblock \bibinfo{title}{{Two techniques for temporal pulse compression in
  gas-filled hollow-core kagomé photonic crystal fiber}}.
\newblock \emph{\bibinfo{journal}{Opt. Lett.}} \textbf{\bibinfo{volume}{38}},
  \bibinfo{pages}{3592--3595} (\bibinfo{year}{2013}).
\newblock
  \urlprefix\url{https://opg.optica.org/ol/abstract.cfm?URI=ol-38-18-3592}.

\bibitem{Schulte2016}
\bibinfo{author}{Schulte, J.}, \bibinfo{author}{Sartorius, T.},
  \bibinfo{author}{Weitenberg, J.}, \bibinfo{author}{Vernaleken, A.} \&
  \bibinfo{author}{Russbueldt, P.}
\newblock \bibinfo{title}{Nonlinear pulse compression in a multi-pass cell}.
\newblock \emph{\bibinfo{journal}{Opt. Lett.}} \textbf{\bibinfo{volume}{41}},
  \bibinfo{pages}{4511--4514} (\bibinfo{year}{2016}).
\newblock
  \urlprefix\url{https://opg.optica.org/ol/abstract.cfm?URI=ol-41-19-4511}.

\bibitem{Weitenberg2017}
\bibinfo{author}{Weitenberg, J.} \emph{et~al.}
\newblock \bibinfo{title}{Multi-pass-cell-based nonlinear pulse compression to
  115 \si{\femto\second} at 7.5 \si{\micro\joule} pulse energy and 300
  \si{\watt} average power}.
\newblock \emph{\bibinfo{journal}{Opt. Express}} \textbf{\bibinfo{volume}{25}},
  \bibinfo{pages}{20502--20510} (\bibinfo{year}{2017}).
\newblock
  \urlprefix\url{https://opg.optica.org/oe/abstract.cfm?URI=oe-25-17-20502}.

\bibitem{MPCReview}
\bibinfo{author}{Viotti, A.-L.} \emph{et~al.}
\newblock \bibinfo{title}{Multi-pass cells for post-compression of ultrashort
  laser pulses}.
\newblock \emph{\bibinfo{journal}{Optica}} \textbf{\bibinfo{volume}{9}},
  \bibinfo{pages}{197--216} (\bibinfo{year}{2022}).
\newblock
  \urlprefix\url{https://opg.optica.org/optica/abstract.cfm?URI=optica-9-2-197}.

\bibitem{TheoryMPC}
\bibinfo{author}{Hanna, M.} \emph{et~al.}
\newblock \bibinfo{title}{Nonlinear temporal compression in multipass cells:
  theory}.
\newblock \emph{\bibinfo{journal}{J. Opt. Soc. Am. B}}
  \textbf{\bibinfo{volume}{34}}, \bibinfo{pages}{1340--1347}
  (\bibinfo{year}{2017}).
\newblock
  \urlprefix\url{https://opg.optica.org/josab/abstract.cfm?URI=josab-34-7-1340}.

\bibitem{SubnsSBSMPC}
\bibinfo{author}{Tarasov, A.~A.} \& \bibinfo{author}{Chu, H.}
\newblock \bibinfo{title}{{Subnanosecond Nd:YAG laser with multipass cell for
  SBS pulse compression}}.
\newblock In \bibinfo{editor}{Clarkson, W.~A.} \& \bibinfo{editor}{Shori,
  R.~K.} (eds.) \emph{\bibinfo{booktitle}{Solid State Lasers XXVI: Technology
  and Devices}}, vol. \bibinfo{volume}{10082}, \bibinfo{pages}{100820Q}.
  \bibinfo{organization}{International Society for Optics and Photonics}
  (\bibinfo{publisher}{SPIE}, \bibinfo{year}{2017}).
\newblock \urlprefix\url{https://doi.org/10.1117/12.2248355}.

\bibitem{HighESBS}
\bibinfo{author}{Feng, C.}, \bibinfo{author}{Xu, X.} \& \bibinfo{author}{Diels,
  J.-C.}
\newblock \bibinfo{title}{{High-energy sub-phonon lifetime pulse compression by
  stimulated Brillouin scattering in liquids}}.
\newblock \emph{\bibinfo{journal}{Opt. Express}} \textbf{\bibinfo{volume}{25}},
  \bibinfo{pages}{12421--12434} (\bibinfo{year}{2017}).
\newblock
  \urlprefix\url{https://opg.optica.org/oe/abstract.cfm?URI=oe-25-11-12421}.

\bibitem{IntLight}
\bibinfo{author}{Maier, M.} \& \bibinfo{author}{Kaiser, W.}
\newblock \bibinfo{title}{{Intense Light Bursts in the Stimulated Raman
  Effect}}.
\newblock \emph{\bibinfo{journal}{Physical Review Letters}}
  \textbf{\bibinfo{volume}{17}} (\bibinfo{year}{1966}).

\bibitem{Johnson1984}
\bibinfo{author}{Johnson, A.~M.}, \bibinfo{author}{Stolen, R.~H.} \&
  \bibinfo{author}{Simpson, W.~M.}
\newblock \bibinfo{title}{{80× single‐stage compression of frequency doubled
  Nd:yttrium aluminum garnet laser pulses}}.
\newblock \emph{\bibinfo{journal}{Applied Physics Letters}}
  \textbf{\bibinfo{volume}{44}}, \bibinfo{pages}{729--731}
  (\bibinfo{year}{1984}).
\newblock \urlprefix\url{https://doi.org/10.1063/1.94897}.
\newblock
  \eprint{https://pubs.aip.org/aip/apl/article-pdf/44/8/729/18450947/729\_1\_online.pdf}.

\bibitem{FiberComprSoliton}
\bibinfo{author}{Mollenauer, L.~F.}, \bibinfo{author}{Stolen, R.~H.},
  \bibinfo{author}{Gordon, J.~P.} \& \bibinfo{author}{Tomlinson, W.~J.}
\newblock \bibinfo{title}{Extreme picosecond pulse narrowing by means of
  soliton effect in single-mode optical fibers}.
\newblock \emph{\bibinfo{journal}{Opt. Lett.}} \textbf{\bibinfo{volume}{8}},
  \bibinfo{pages}{289--291} (\bibinfo{year}{1983}).
\newblock
  \urlprefix\url{https://opg.optica.org/ol/abstract.cfm?URI=ol-8-5-289}.

\bibitem{Sub40fsMPC}
\bibinfo{author}{Weitenberg, J.}, \bibinfo{author}{Saule, T.},
  \bibinfo{author}{Schulte, J.} \& \bibinfo{author}{Rubbuldt, P.}
\newblock \bibinfo{title}{Nonlinear pulse compression to sub-40
  \si{\femto\second} at 4.5 \si{\micro\joule} pulse energy by multi-pass-cell
  spectral broadening}.
\newblock \emph{\bibinfo{journal}{IEEE Journal of Quantum Electronics}}
  \textbf{\bibinfo{volume}{PP}}, \bibinfo{pages}{1--1} (\bibinfo{year}{2017}).

\bibitem{Gen172fsMPC}
\bibinfo{author}{Song, J.} \emph{et~al.}
\newblock \bibinfo{title}{Generation of 172 \si{\femto\second} pulse from a
  {Nd:YVO4} picosecond laser by using multi‑pass‑cell technique}.
\newblock \emph{\bibinfo{journal}{Applied Physics B}}
  \textbf{\bibinfo{volume}{127}} (\bibinfo{year}{2021}).

\bibitem{FLASHpsMPC}
\bibinfo{author}{Seidel, M.} \emph{et~al.}
\newblock \bibinfo{title}{Ultrafast \si{\mega\hertz}-rate burst-mode pump-probe
  laser for the {FLASH FEL} facility based on nonlinear compression of
  \si{\pico\second}-level pulses from an yb-amplifier chain}.
\newblock \emph{\bibinfo{journal}{Laser \& Photonics Reviews}}
  \textbf{\bibinfo{volume}{16}} (\bibinfo{year}{2022}).

\bibitem{DispEngMPC}
\bibinfo{author}{Silletti, L.} \emph{et~al.}
\newblock \bibinfo{title}{Dispersion-engineered multi-pass cell for
  single-stage post-compression of an ytterbium laser}.
\newblock \emph{\bibinfo{journal}{Opt. Lett.}} \textbf{\bibinfo{volume}{48}},
  \bibinfo{pages}{1842--1845} (\bibinfo{year}{2023}).
\newblock
  \urlprefix\url{https://opg.optica.org/ol/abstract.cfm?URI=ol-48-7-1842}.

\bibitem{HighEGasMPC}
\bibinfo{author}{Ueffing, M.} \emph{et~al.}
\newblock \bibinfo{title}{Nonlinear pulse compression in a gas-filled multipass
  cell}.
\newblock \emph{\bibinfo{journal}{Opt. Lett.}} \textbf{\bibinfo{volume}{43}},
  \bibinfo{pages}{2070--2073} (\bibinfo{year}{2018}).
\newblock
  \urlprefix\url{https://opg.optica.org/ol/abstract.cfm?URI=ol-43-9-2070}.

\bibitem{FELMPC}
\bibinfo{author}{Viotti, A.~L.} \emph{et~al.}
\newblock \bibinfo{title}{60 \si{\femto\second}, 1030 \si{\nano\meter} {FEL}
  pump–probe laser based on a multi-pass post-compressed {Yb:YAG} source}.
\newblock \emph{\bibinfo{journal}{Journal of Synchrotron Radiation}}
  \textbf{\bibinfo{volume}{28}} (\bibinfo{year}{2021}).

\bibitem{LG10MPC}
\bibinfo{author}{Kaumanns, M.}, \bibinfo{author}{Kormin, D.},
  \bibinfo{author}{Nubbemeyer, T.}, \bibinfo{author}{Pervak, V.} \&
  \bibinfo{author}{Karsch, S.}
\newblock \bibinfo{title}{Spectral broadening of 112\si{mJ}, 1.3\si{ps} pulses
  at 5\si{kHz} in a {LG10} multipass cell with compressibility to 37\si{fs}}.
\newblock \emph{\bibinfo{journal}{Opt. Lett.}} \textbf{\bibinfo{volume}{46}},
  \bibinfo{pages}{929--932} (\bibinfo{year}{2021}).
\newblock
  \urlprefix\url{https://opg.optica.org/ol/abstract.cfm?URI=ol-46-5-929}.

\bibitem{Gen300psSBS}
\bibinfo{author}{Feng, C.}, \bibinfo{author}{Xu, X.} \& \bibinfo{author}{Diels,
  J.-C.}
\newblock \bibinfo{title}{Generation of 300 \si{\pico\second} laser pulse with
  1.2 \si{\joule} energy by stimulated {Brillouin} scattering in water at 532
  \si{\nano\meter}}.
\newblock \emph{\bibinfo{journal}{CLEO: 2014}} \bibinfo{pages}{SM2F.7}
  (\bibinfo{year}{2014}).
\newblock
  \urlprefix\url{https://opg.optica.org/abstract.cfm?URI=CLEO_SI-2014-SM2F.7}.

\bibitem{Kafka1984}
\bibinfo{author}{Kafka, J.~D.}, \bibinfo{author}{Kolner, B.~H.},
  \bibinfo{author}{Baer, T.} \& \bibinfo{author}{Bloom, D.~M.}
\newblock \bibinfo{title}{Compression of pulses from a continuous-wave
  mode-locked nd:yag laser}.
\newblock \emph{\bibinfo{journal}{Optics Letters}}
  \textbf{\bibinfo{volume}{9}}, \bibinfo{pages}{505} (\bibinfo{year}{1984}).

\bibitem{Niemz1995}
\bibinfo{author}{Niemz, M.~H.}
\newblock \bibinfo{title}{Threshold dependence of laser-induced optical
  breakdown on pulse duration}.
\newblock \emph{\bibinfo{journal}{Applied Physics Letters}}
  \textbf{\bibinfo{volume}{66}}, \bibinfo{pages}{1181--1183}
  (\bibinfo{year}{1995}).

\bibitem{Atkocaitis2022}
\bibinfo{author}{Atkočaitis, E.}, \bibinfo{author}{Smalakys, L.} \&
  \bibinfo{author}{Melninkaitis, A.}
\newblock \bibinfo{title}{Pulse temporal scaling of lidt for anti-reflective
  coatings deposited on lithium triborate crystals}.
\newblock \emph{\bibinfo{journal}{Optics Express}}
  \textbf{\bibinfo{volume}{30}}, \bibinfo{pages}{28401} (\bibinfo{year}{2022}).

\bibitem{Couairon2011}
\bibinfo{author}{Couairon, A.} \emph{et~al.}
\newblock \bibinfo{title}{Practitioner’s guide to laser pulse propagation
  models and simulation}.
\newblock \emph{\bibinfo{journal}{The European Physical Journal Special
  Topics}} \textbf{\bibinfo{volume}{199}} (\bibinfo{year}{2011}).

\bibitem{Robert2007}
\bibinfo{author}{Robert, C.}
\newblock \bibinfo{title}{Simple, stable, and compact multiple-reflection
  optical cell for very long optical paths}.
\newblock \emph{\bibinfo{journal}{Appl. Opt.}} \textbf{\bibinfo{volume}{46}},
  \bibinfo{pages}{5408--5418} (\bibinfo{year}{2007}).
\newblock
  \urlprefix\url{https://opg.optica.org/ao/abstract.cfm?URI=ao-46-22-5408}.

\bibitem{Glebov2014}
\bibinfo{author}{Glebov, L.} \emph{et~al.}
\newblock \bibinfo{title}{Volume-chirped bragg gratings: monolithic components
  for stretching and compression of ultrashort laser pulses}.
\newblock \emph{\bibinfo{journal}{Optical Engineering}}
  \textbf{\bibinfo{volume}{53}}, \bibinfo{pages}{051514}
  (\bibinfo{year}{2014}).

\bibitem{Corless1997}
\bibinfo{author}{Corless, J.~D.}, \bibinfo{author}{West, J.~A.},
  \bibinfo{author}{Bromage, J.} \& \bibinfo{author}{Stroud, C.~R.}
\newblock \bibinfo{title}{Pulsed single-mode dye laser for coherent control
  experiments}.
\newblock \emph{\bibinfo{journal}{Review of Scientific Instruments}}
  \textbf{\bibinfo{volume}{68}}, \bibinfo{pages}{2259--2264}
  (\bibinfo{year}{1997}).

\bibitem{Chiao1964}
\bibinfo{author}{Chiao, R.~Y.}, \bibinfo{author}{Townes, C.~H.} \&
  \bibinfo{author}{Stoicheff, B.~P.}
\newblock \bibinfo{title}{{Stimulated Brillouin Scattering and Coherent
  Generation of Intense Hypersonic Waves}}.
\newblock \emph{\bibinfo{journal}{Phys. Rev. Lett.}}
  \textbf{\bibinfo{volume}{12}}, \bibinfo{pages}{592--595}
  (\bibinfo{year}{1964}).
\newblock \urlprefix\url{https://link.aps.org/doi/10.1103/PhysRevLett.12.592}.

\bibitem{Armandillo1983}
\bibinfo{author}{Armandillo, E.} \& \bibinfo{author}{Proch, D.}
\newblock \bibinfo{title}{{Highly efficient, high-quality phase-conjugate
  reflection at 308 nm using stimulated Brillouin scattering}}.
\newblock \emph{\bibinfo{journal}{Opt. Lett.}} \textbf{\bibinfo{volume}{8}},
  \bibinfo{pages}{523--525} (\bibinfo{year}{1983}).
\newblock
  \urlprefix\url{https://opg.optica.org/ol/abstract.cfm?URI=ol-8-10-523}.

\bibitem{Agrawal2013}
\bibinfo{author}{Agrawal, G.~P.}
\newblock \emph{\bibinfo{title}{Nonlinear fiber optics}}.
\newblock Engineering professional collection (\bibinfo{publisher}{Academic
  Press}, \bibinfo{address}{Oxford}, \bibinfo{year}{2013}),
  \bibinfo{edition}{5th ed} edn.
\newblock \bibinfo{note}{Includes bibliographical references and index}.

\bibitem{Hanna2020}
\bibinfo{author}{Hanna, M.}, \bibinfo{author}{Daher, N.},
  \bibinfo{author}{Guichard, F.}, \bibinfo{author}{D\'{e}len, X.} \&
  \bibinfo{author}{Georges, P.}
\newblock \bibinfo{title}{Hybrid pulse propagation model and
  quasi-phase-matched four-wave mixing in multipass cells}.
\newblock \emph{\bibinfo{journal}{J. Opt. Soc. Am. B}}
  \textbf{\bibinfo{volume}{37}}, \bibinfo{pages}{2982--2988}
  (\bibinfo{year}{2020}).
\newblock
  \urlprefix\url{https://opg.optica.org/josab/abstract.cfm?URI=josab-37-10-2982}.

\bibitem{Escoto2022}
\bibinfo{author}{Escoto, E.} \emph{et~al.}
\newblock \bibinfo{title}{Temporal quality of post-compressed pulses at large
  compression factors}.
\newblock \emph{\bibinfo{journal}{J. Opt. Soc. Am. B}}
  \textbf{\bibinfo{volume}{39}}, \bibinfo{pages}{1694--1702}
  (\bibinfo{year}{2022}).
\newblock
  \urlprefix\url{https://opg.optica.org/josab/abstract.cfm?URI=josab-39-7-1694}.

\bibitem{Escoto2024}
\bibinfo{author}{Escoto, E.} \emph{et~al.}
\newblock \bibinfo{title}{Improved temporal characteristics for post-compressed
  pulses via application-tailored nonlinear polarization ellipse rotation}.
\newblock \emph{\bibinfo{journal}{Opt. Lett.}} \textbf{\bibinfo{volume}{49}},
  \bibinfo{pages}{6841--6844} (\bibinfo{year}{2024}).
\newblock
  \urlprefix\url{https://opg.optica.org/ol/abstract.cfm?URI=ol-49-23-6841}.

\bibitem{Pfaff2022}
\bibinfo{author}{Pfaff, Y.} \emph{et~al.}
\newblock \bibinfo{title}{Nonlinear pulse compression of a thin-disk amplifier
  and contrast enhancement via nonlinear ellipse rotation}.
\newblock \emph{\bibinfo{journal}{Opt. Express}} \textbf{\bibinfo{volume}{30}},
  \bibinfo{pages}{10981--10990} (\bibinfo{year}{2022}).
\newblock
  \urlprefix\url{https://opg.optica.org/oe/abstract.cfm?URI=oe-30-7-10981}.

\bibitem{Benner2023}
\bibinfo{author}{Benner, M.}, \bibinfo{author}{Karst, M.},
  \bibinfo{author}{Mendez, C.~A.}, \bibinfo{author}{Stark, H.} \&
  \bibinfo{author}{Limpert, J.}
\newblock \bibinfo{title}{Concept of enhanced frequency chirping for multi-pass
  cells to improve the pulse contrast}.
\newblock \emph{\bibinfo{journal}{J. Opt. Soc. Am. B}}
  \textbf{\bibinfo{volume}{40}}, \bibinfo{pages}{301--305}
  (\bibinfo{year}{2023}).
\newblock
  \urlprefix\url{https://opg.optica.org/josab/abstract.cfm?URI=josab-40-2-301}.

\bibitem{Nguyen2011}
\bibinfo{author}{Nguyen, D.}, \bibinfo{author}{Piracha, M.~U.},
  \bibinfo{author}{Mandridis, D.} \& \bibinfo{author}{Delfyett, P.~J.}
\newblock \bibinfo{title}{Dynamic parabolic pulse generation using temporal
  shaping of wavelength to time mapped pulses}.
\newblock \emph{\bibinfo{journal}{Opt. Express}} \textbf{\bibinfo{volume}{19}},
  \bibinfo{pages}{12305--12311} (\bibinfo{year}{2011}).
\newblock
  \urlprefix\url{https://opg.optica.org/oe/abstract.cfm?URI=oe-19-13-12305}.

\bibitem{Rogers2016}
\bibinfo{author}{Rogers, C.~E.} \& \bibinfo{author}{Gould, P.~L.}
\newblock \bibinfo{title}{Nanosecond pulse shaping at 780 nm with fiber-based
  electro-optical modulators and a double-pass tapered amplifier}.
\newblock \emph{\bibinfo{journal}{Opt. Express}} \textbf{\bibinfo{volume}{24}},
  \bibinfo{pages}{2596--2606} (\bibinfo{year}{2016}).
\newblock
  \urlprefix\url{https://opg.optica.org/oe/abstract.cfm?URI=oe-24-3-2596}.

\bibitem{Meijer2017}
\bibinfo{author}{Meijer, R.~A.}, \bibinfo{author}{Stodolna, A.~S.},
  \bibinfo{author}{Eikema, K. S.~E.} \& \bibinfo{author}{Witte, S.}
\newblock \bibinfo{title}{High-energy {Nd:YAG} laser system with arbitrary
  sub-nanosecond pulse shaping capability}.
\newblock \emph{\bibinfo{journal}{Opt. Lett.}} \textbf{\bibinfo{volume}{42}},
  \bibinfo{pages}{2758--2761} (\bibinfo{year}{2017}).
\newblock
  \urlprefix\url{https://opg.optica.org/ol/abstract.cfm?URI=ol-42-14-2758}.

\bibitem{Sooy1965}
\bibinfo{author}{Sooy, W.~R.}
\newblock \bibinfo{title}{The natural selection of modes in a passive
  {Q}‐switched laser}.
\newblock \emph{\bibinfo{journal}{Applied Physics Letters}}
  \textbf{\bibinfo{volume}{7}}, \bibinfo{pages}{36--37} (\bibinfo{year}{1965}).
\newblock \urlprefix\url{https://doi.org/10.1063/1.1754286}.
\newblock
  \eprint{https://pubs.aip.org/aip/apl/article-pdf/7/2/36/18417952/36\_1\_online.pdf}.

\bibitem{Bai2018}
\bibinfo{author}{Bai, Z.} \emph{et~al.}
\newblock \bibinfo{title}{{Stimulated Brillouin scattering materials,
  experimental design and applications: A review}}.
\newblock \emph{\bibinfo{journal}{Optical Materials}}
  \textbf{\bibinfo{volume}{75}}, \bibinfo{pages}{626--645}
  (\bibinfo{year}{2018}).
\newblock
  \urlprefix\url{https://www.sciencedirect.com/science/article/pii/S0925346717306687}.

\bibitem{Keaton2014}
\bibinfo{author}{Keaton, G.~L.}, \bibinfo{author}{Leonardo, M.~J.},
  \bibinfo{author}{Byer, M.~W.} \& \bibinfo{author}{Richard, D.~J.}
\newblock \bibinfo{title}{{Stimulated Brillouin scattering of pulses in optical
  fibers}}.
\newblock \emph{\bibinfo{journal}{Opt. Express}} \textbf{\bibinfo{volume}{22}},
  \bibinfo{pages}{13351--13365} (\bibinfo{year}{2014}).
\newblock
  \urlprefix\url{https://opg.optica.org/oe/abstract.cfm?URI=oe-22-11-13351}.

\bibitem{Stolen1989}
\bibinfo{author}{Stolen, R.~H.}, \bibinfo{author}{Tomlinson, W.~J.},
  \bibinfo{author}{Haus, H.~A.} \& \bibinfo{author}{Gordon, J.~P.}
\newblock \bibinfo{title}{{Raman response function of silica-core fibers}}.
\newblock \emph{\bibinfo{journal}{Journal of the Optical Society of America B}}
  \textbf{\bibinfo{volume}{6}}, \bibinfo{pages}{1159} (\bibinfo{year}{1989}).

\bibitem{Nakashima1987}
\bibinfo{author}{Nakashima, T.}, \bibinfo{author}{Nakazawa, M.},
  \bibinfo{author}{Nishi, K.} \& \bibinfo{author}{Kubota, H.}
\newblock \bibinfo{title}{{Effect of stimulated Raman scattering on
  pulse-compression characteristics}}.
\newblock \emph{\bibinfo{journal}{Opt. Lett.}} \textbf{\bibinfo{volume}{12}},
  \bibinfo{pages}{404--406} (\bibinfo{year}{1987}).
\newblock
  \urlprefix\url{https://opg.optica.org/ol/abstract.cfm?URI=ol-12-6-404}.

\bibitem{Chen2023}
\bibinfo{author}{Chen, B.} \emph{et~al.}
\newblock \bibinfo{title}{{Gain characteristics of stimulated Brillouin
  scattering in fused silica}}.
\newblock \emph{\bibinfo{journal}{Opt. Express}} \textbf{\bibinfo{volume}{31}},
  \bibinfo{pages}{5699--5707} (\bibinfo{year}{2023}).
\newblock
  \urlprefix\url{https://opg.optica.org/oe/abstract.cfm?URI=oe-31-4-5699}.

\bibitem{Boyd2020}
\bibinfo{author}{Boyd, R.~W.}
\newblock \emph{\bibinfo{title}{Nonlinear optics}}
  (\bibinfo{publisher}{Elsevier, AP Academic Press}, \bibinfo{address}{London},
  \bibinfo{year}{2020}), \bibinfo{edition}{fourth edition} edn.

\bibitem{Kabacinski:19}
\bibinfo{author}{Kabaci\'{n}ski, P.}, \bibinfo{author}{Karda\'{s}, T.~M.},
  \bibinfo{author}{Stepanenko, Y.} \& \bibinfo{author}{Radzewicz, C.}
\newblock \bibinfo{title}{Nonlinear refractive index measurement by
  {SPM}-induced phase regression}.
\newblock \emph{\bibinfo{journal}{Opt. Express}} \textbf{\bibinfo{volume}{27}},
  \bibinfo{pages}{11018--11028} (\bibinfo{year}{2019}).
\newblock
  \urlprefix\url{https://opg.optica.org/oe/abstract.cfm?URI=oe-27-8-11018}.

\bibitem{Paschotta_2005_dielectric_mirrors}
\bibinfo{author}{Paschotta, R.}
\newblock \bibinfo{title}{Dielectric mirrors}.
\newblock \bibinfo{howpublished}{RP Photonics Encyclopedia}
  (\bibinfo{year}{2005}).
\newblock \urlprefix\url{https://www.rp-photonics.com/dielectric_mirrors.html}.
\newblock \bibinfo{note}{Available online at
  \url{https://www.rp-photonics.com/dielectric_mirrors.html}}.

\bibitem{Headley1996}
\bibinfo{author}{Headley, C.} \& \bibinfo{author}{Agrawal, G.~P.}
\newblock \bibinfo{title}{{Unified description of ultrafast stimulated Raman
  scattering in optical fibers}}.
\newblock \emph{\bibinfo{journal}{Journal of the Optical Society of America B}}
  \textbf{\bibinfo{volume}{13}}, \bibinfo{pages}{2170} (\bibinfo{year}{1996}).

\bibitem{Malitson1965}
\bibinfo{author}{Malitson, I.~H.}
\newblock \bibinfo{title}{Interspecimen comparison of the refractive index of
  fused silica*,†}.
\newblock \emph{\bibinfo{journal}{Journal of the Optical Society of America}}
  \textbf{\bibinfo{volume}{55}}, \bibinfo{pages}{1205} (\bibinfo{year}{1965}).

\bibitem{Kabacinski2019}
\bibinfo{author}{Kabaci\'{n}ski, P.}, \bibinfo{author}{Karda\'{s}, T.~M.},
  \bibinfo{author}{Stepanenko, Y.} \& \bibinfo{author}{Radzewicz, C.}
\newblock \bibinfo{title}{Nonlinear refractive index measurement by spm-induced
  phase regression}.
\newblock \emph{\bibinfo{journal}{Opt. Express}} \textbf{\bibinfo{volume}{27}},
  \bibinfo{pages}{11018--11028} (\bibinfo{year}{2019}).
\newblock
  \urlprefix\url{https://opg.optica.org/oe/abstract.cfm?URI=oe-27-8-11018}.

\end{thebibliography}

\end{document}